\documentclass[preprint,preprintnumbers,amsmath,amssymb]{revtex4}
\usepackage{amsfonts}
\usepackage{amssymb}
\usepackage{latexsym}

\usepackage{lscape}

\newcommand{\al}{\alpha}

\newcommand{\pa}{\partial}

\newcommand{\la}{\lambda}
\newcommand{\La}{\Lambda}

\newcommand{\ga}{\gamma}
\newcommand{\om}{\omega}

\newcommand{\non}{\nonumber}

\begin{document}

\begin{titlepage}

\begin{center}

{\Large An infinite family of solvable and integrable quantum systems on a plane}

\vskip 0.7cm

Fr\'ed\'erick Tremblay,\\
Centre de recherches math\'ematiques and D\'epartement de math\'ematiques et de statistique,\\
Universit\'e de Montreal,  C.P. 6128,
succ. Centre-ville, Montr\'eal (QC) H3C 3J7, Canada, \\[10pt]

Alexander V. Turbiner,\\
Instituto de Ciencias Nucleares, UNAM, A.P. 70-543, 04510 M\'exico D.F., Mexico,\\[10pt]

 and\\[7pt]

Pavel Winternitz, \\
Centre de recherches math\'ematiques and D\'epartement de math\'ematiques et de statistique,\\
Universit\'e de Montreal,  C.P. 6128,
succ. Centre-ville, Montr\'eal (QC) H3C 3J7, Canada
\end{center}

\vskip .9cm

\centerline{\large Abstract}

An infinite family of exactly-solvable and integrable potentials on a plane
is introduced. It is shown that all already known rational potentials with the above properties allowing separation of variables in polar coordinates are particular
cases of this family. The underlying algebraic structure
of the new potentials is revealed as well as its hidden algebra. We conjecture that
all members of the family are also superintegrable and demonstrate this for the
first few cases. A quasi-exactly-solvable and integrable generalization of the family is found.

\end{titlepage}

\section{Introduction}

Some quantum mechanical systems can be characterized by two differently
defined global properties. The first has been called exact solvability
and it means that all energy levels can be calculated algebraically
and the corresponding  wave functions can be obtained as polynomials
in the appropriate variables, multiplied by some overall gauge factor.
The other property is that of integrability, namely the existence of
$n$ integrals of motion that are well defined quantum mechanical
operators, commuting with the Hamiltonian and amongst each other.

A more restrictive property than integrability is superintegrability: the existence of more integrals of motion than degrees of freedom. A maximally superintegrable system has $2n-1$ integrals of motion, including the Hamiltonian. Only subsets of $n$ of them can commute amongst each other.

Any one-dimensional system is integrable and aslo maximally superintegrable by definition. In this article we concentrate on the two-dimensional case where the situation is quite different. A two-dimensional system is integrable if it allows two integrals of motion and maximally superintegrable if it allows three. Some time ago, it was conjectured  that all maximally superintegrable systems for $n=2$ are exactly solvable \cite{TTW:2001}. Here we will show that several exactly-solvable systems are in fact maximally superintegrable. Though they seem very different, they are particular cases of a parametric family of Hamiltonians.

Let us consider the following Hamiltonian in ${\bf R^2}$ written in polar coordinates

\begin{equation}
\label{H}
 H_k (r,\varphi;\om, \al, \beta)\ =\ -\pa_r^2 - \frac{1}{r}\pa_r - \frac{1}{r^2}\pa_{\varphi}^2  + \om^2 r^2 + \frac{\al k^2}{r^2 \cos^2 {k \varphi}} + \frac{\beta k^2}{r^2 \sin^2 {k \varphi}}\ ,
\end{equation}
where $\al, \beta > - \frac{1}{4 k^2}$, $\om$ and $k \neq 0$ are parameters. For $k=1$ this system was introduced in  \cite{Fris:1965, Wint:1966} and has been called the Smorodinsky-Winternitz system \cite{evans:1990}. For $k=2$ the Hamiltonian (1) corresponds to the so-called rational $BC_2$ model \cite{Olshanetsky:1983, Turbiner:1998}. For $k=3$ it describes the Wolfes model \cite{Wolfes:1974}
(it is the rational $G_2$ model in the Hamiltonian reduction method nomenclature \cite{Olshanetsky:1983, Turbiner:1998}); if $\al=0$ it reduces to the Calogero model \cite{Calogero}. The configuration space of (\ref{H}) is given by the sector $\frac{\pi}{2 k} \geq \varphi \geq 0\ ,\ r \in [0,\infty)$  which is  the Weyl chamber for $BC_2$ if $k=2$ and $G_2$ if $k=3$, respectively.

There is an interesting feature of the Hamiltonian (\ref{H}) connecting different values of $k$, namely,
\begin{equation}
\label{fea1}
 H_{2 \ell} (r,\varphi;\om, 0, \beta)\ =\ H_{\ell} (r,\varphi;\om, \beta, \beta)\ ,
\end{equation}
\begin{equation}
\label{fea2}
 H_{2 \ell} (r,\varphi;\om, \al, 0)\ =\ H_{\ell} (r,\varphi -\frac{\pi}{4\ell };\om, \al, \al)\  .
\end{equation}

The Hamiltonian (\ref{H}) to our knowledge includes {\it all} published superintegrable systems in a Euclidian plane $E_2$ that allow the separation of variables in polar coordinates.

\section{Exact-solvability}
It is well-known that the model (\ref{H}) for $k=1,2,3$ is exactly-solvable (the energies and eigenfunctions can be found explicitly). For $\alpha=0$ and $k$ integer the exact solvability of the Hamiltonian (1) was mentioned in Ref.\cite{Khare}.
It can be immediately checked by a direct calculation that the ground state of (\ref{H}) is given by
\begin{equation}
\label{GS}
 \Psi_0 \ =\ r^{(a+b)k}\ \cos^a {k \varphi}\ \sin^b {k \varphi}\ e^{-\frac{\om r^2}{2}}\ , \ E_0\ =\ 2\om [(a + b)k +1]\ ,
\end{equation}
where $\al=a(a-1)$ and $\beta=b(b-1)$\ . If we make a gauge rotation of the Hamiltonian (\ref{H}),
\begin{equation}
 h_k\ =\ \Psi_0^{-1} (H_k - E_0) \Psi_0 \,,
\end{equation}
we obtain the operator
\begin{equation}
\label{h}
 h_k \ =\ -\pa_r^2 +(2\om r - \frac{2k(a+b) + 1}{r})\pa_r - \frac{1}{r^2}\pa_{\varphi}^2
 - \frac{2k}{r^2}(-a\tan {k\varphi}+b\cot {k\varphi})\pa_{\varphi}\ ,
\end{equation}
for which the lowest eigenfunction is a constant with zero eigenvalue.

The original eigenfunctions $\Psi(r, \varphi)$  of the Hamiltonian (\ref{H}) are related to those of the transformed Hamiltonian $h_k$ as follows $\Psi(r, \varphi)=\Psi_0(r, \varphi) \Xi (r, \varphi)$.
Let us solve the original problem (\ref{H}) in a traditional way by a separation of variables in $h_k$. Thus we  assume  $\Xi(r,\varphi)= R(r) \Phi(\varphi)$ and write
\begin{equation}
\label{h_k}
  h_k\ =\ h_r + \frac{1}{r^2} h_{\varphi}\ .
\end{equation}
The operator $h_{\varphi}$ written in the new coordinate $z = {\sin}^2 {k \varphi}$ reads
\begin{equation}
\label{h_phi}
  h_{\varphi}\ =\ 4k^2 z (z-1) \pa_z^2 + 4 k^2 [(a+b+1)z-b-\frac{1}{2}]
  \pa_z   \ .
\end{equation}
The eigenvalue problem $h_{\varphi}\Phi \ =\ \La_n \Phi$ where $\La_n$ is the separation constant has polynomial eigenfunctions
\begin{equation}
\label{h_Phi_n}
 \Phi_n (z) = P_n^{(a-1/2,b-1/2)}(2z-1)\ ,\ \La_n\ =\ 4k^2 n (n+a+b)\ ,\ n=0,1,2,\ldots
\end{equation}
where $P_n^{(a-1/2,b-1/2)}(2z-1)$ is a Jacobi polynomial. Now the eigenvalue problem for the operator
\begin{equation}
\label{h_k1}
 h_k \ =\ -\pa_r^2 +(2\om r- \frac{2ak +2bk + 1}{r})\pa_r + \frac{\La_n}{r^2} \ ,
\end{equation}
appears. Let us perform a further gauge rotation of $h_r$,
\begin{equation}
\label{h_kt}
 \tilde h_k = r^{-\ga} h_k r^{\ga} = -\pa_r^2 +(2\om r- \frac{2ak +2bk + 2 \ga + 1}{r})\pa_r + \frac{\La_n - 2 (a+b) k \ga - \ga^2}{r^2}
 + 2\om \ga \ .
\end{equation}
We absorb the term $2\om \ga$ in the energy and choose $\ga = 2 k n$ so as to remove the $1/r^2$ term,
\[
  \ga^2 + 2 (a+b) k \ga - \La_n = 0 \ .
\]

The resulting radial operator in the $t=r^2$ variable
\begin{equation}
\label{h_r}
 \tilde h_r\ =\ -4t \pa_t^2 + 4[\om t - k(2n+ a + b) - 1]\pa_t \ ,
\end{equation}
has the eigenstates
\begin{equation}
\label{R_N}
 R_N(t)\ =\ L_N^{(k(2n+ a + b))} (\om t)\ ,\ E_N=4\om N\ ,
\end{equation}
where $L_N^{(k(2n+ a + b))} (\om t)$ is a Laguerre polynomial. Finally, the eigenstates of (\ref{H}) are
\begin{equation}
\label{Psi_N}
  \Psi_{N,n}\ =\ r^{2 n k} R_N(r^2) P_n^{(a-1/2 , b-1/2)}(2{\sin}^2 {k \varphi}-1) \Psi_0 \ ,\ E_{N,n}\ =\ 2\om [2N+(2n + a+b)k+1]\ .
\end{equation}
All formulas remain valid for any real $k \neq 0$. In particular, both $R_N(r^2)$ and $P_n^{(\al,\beta)}(z)$ remain polynomials.
The eigenvalues are linear in the quantum numbers $N,n$. For integer (and rational) $k$ there is a degeneracy of states, which is determined by the  number of solutions of the equation
\[
  N+kn = \mbox{integer}\ .
\]
Varying $k$ we can change degeneracy leaving the spectra linear in the quantum numbers $N, n$.

If $k$ in (\ref{H}) takes integer values we have a Lie-algebraic interpretation of the problem (\ref{H}). In order to reveal it let us make the following change of variables,
\begin{equation}
\label{coord}
t\ =\ r^2\ ,\ u\ =\ r^{2k} \sin^2 {k\varphi}\ .
\end{equation}
The resulting gauge-transformed Hamiltonian (\ref{h}) in these coordinates takes an algebraic form:
\[
  h_k\ =   -4 t \pa^2_t - 8k u \pa^2_{tu} - 4k^2 t^{k-1} u \pa^2_u
\]
\begin{equation}
\label{Lie}
    + 4[\om t - (a+b)k -1]\pa_t + [4\om k u - 2 k^2(2 b + 1)
    t^{k-1}]\pa_u \ .
\end{equation}
It coincides with already known expressions for the Hamiltonian for $k=1,2,3$ in appropriate variables (see \cite{TTW:2001, Turbiner:1998}).
What is its underlying hidden algebra if any?

The Hamiltonian $h_k$ preserves the space of polynomials
\begin{equation}
\label{space_r}
 {\cal P}_{{\cal N}}^{(s)}\ =\ (t^{p} u^{q} | 0\leq (p + s q) \leq {\cal N})\ , \quad {\cal N} = 0,1,2,\ldots  \ ,
\end{equation}
for $s \geq k-1$ and any integer ${\cal N}$. Hence, it has infinitely many finite-dimensional invariant subspaces ${\cal P}_{{\cal N}}^{(s)}$. These spaces can be ordered forming an {\it infinite flag},
\begin{equation}
\label{flag}
  {\cal P}_0^{(s)} \subset {\cal P}_1^{(s)} \subset {\cal P}_2^{(s)} \ldots {\cal P}_{\cal N}^{(s)} \ldots
\end{equation}
for fixed $s$. We call this flag  ${\cal P}^{(s)}$.
The space ${\cal P}_{{\cal N}}^{(s)}$ is a finite-dimensional irreducible representation space of the infinite-dimensional finitely-generated Lie algebra $g^{(s)} \supset gl(2,{\bf R}) \ltimes {\bf R}^{s+1} \oplus T_s$ of monomials in $(s+6)-$ generating operators. These generating operators are \cite{gko1} (see also \cite{Turbiner:1994} and \cite{ghko}),
\[ J^1\  =\  \pa_t \ , \]
\[ J^2_{\cal N}\  =\ t \pa_t\ -\ \frac{{\cal N}}{3} \ ,
 \ J^3_{\cal N}\  =\ s u\pa_u\ -\ \frac{{\cal N}}{3}\ ,\]
\[ J^4_{\cal N}\  =\ t^2 \pa_t \  +\ s t u \pa_u \ - \ {\cal N} t \ ,\]
\begin{equation}
\label{gr}
 R_{i}\  = \ t^{i}\pa_u\ ,\quad i=0,1,\dots, s\ ,
\end{equation}
and
\begin{equation}
\label{grT}
 T_s\ =\ u\pa_{t}^s\ ,
\end{equation}
(see \cite{Turbiner:1998}).
The generator $J^3_{\cal N}$ is the central generator of the $gl(2,{\bf R})$-algebra. The generators (19) of the non-semi-simple Lie algebra $gl(2) \ltimes R^{s+1}$ are vector fields on line bundles over a s-Hirzebruch surface \cite{ghko}. The meaning of the generator (\ref{grT}) for $s>1$ is unclear.

For $s=1$ the algebra $g^{(1)}$ coincides with the  algebra $sl(3)$. It has the space (\ref{space_r}) for $s=1$ as an invariant subspace and acts irreducibly there. It is important to note that the space ${\cal P}_{{\cal N}}^{(g)}$ is a finite-dimensional (reducible) representation space of the finite-dimensional non-semisimple Lie algebra $gl(2,{\bf R}) \ltimes {\bf R}^{g+1}$ (see \cite{Turbiner:1998}),
\begin{equation}
\label{space_r,p}
 \tilde {\cal P}_{{\cal N},p}^{(s)}\ =\
 \langle t^{n_1} u^{n_2} | 0\leq (n_1+s n_2)
\leq {\cal N} \quad \mbox{and}\quad 0\leq n_2 \leq p \rangle
\end{equation}
For fixed $s$ and $p$ these spaces form the flag $\tilde {\cal P}_{p}^{(s)}$. Each such a flag for $s \geq k-1$ is preserved by the Hamiltonian $h_k$. This gives information about the structure of the eigenfunctions. In particular, it implies the existence of a family of eigenfunctions which depend on the variable $t$ only.

It can be immediately checked that $h_k$ for fixed integer $k$ preserves the flag ${\cal P}^{(s)}$ for $s=k-1$, $s=k$, or $s > k$ assuming the hidden algebras $g^{(k-1)}$, $g^{(k)}$, $g^{(s)}$, respectively. It is worth mentioning that the first case $s=k-1$ supports the already known hidden algebras of the trigonometric $BC_2$ for $k=2$ and $G_2$ for $k=3$ models, respectively, contrary to the second or third case. However, later on we will see that the case $s=k-1$ is excluded (see Section III).

The fact that $s$ can take any values $s \geq k-1$ reflects a degeneracy of eigenstates of the original problem (\ref{H}). For particular cases $k=2,3$ it was already mentioned in the paper \cite{Turbiner:1998}. This degeneracy is removed by the algebraic form of the integrals of motion (see below). Hence, for any integer $k$ the algebraic Hamiltonian (\ref{Lie}) can be rewritten in terms of the generators (\ref{gr}) (without the operator $J^4_{\cal N}$) (see Theorem 4.3 from \cite{Turbiner:1994}):
\[
  h_k\ =   -4 J^2 J^1 - 8 J^3 J^1 - 4k R_{k-1} J^3
\]
\begin{equation}
\label{Lie_gen}
    + 4\om J^2 - 4[(a+b)k -1]J^1 + 4\om J^3 - 2 k^2(2 b + 1) R_{k-1} \ .
\end{equation}
where $J^i \equiv J^i_{0}$.

\section{\bf Complete Integrability.}

It is obvious that
\begin{equation}
\label{X_k}
   {\cal X}_k (\al, \beta)\ =\ -L_3^2 + \frac{\al k^2}{\cos^2 {k \varphi}} + \frac{\beta k^2}{\sin^2 {k \varphi}}\ ,
\end{equation}
where $L_3 = \pa_{\varphi}$ is the 2D angular momentum, is an integral of motion \cite{Fris:1965, Wint:1966}.
Its existence is directly related to the separation of variables in polar coordinates in the Schroedinger equation for (\ref{H}).
Therefore the Hamiltonian (\ref{H}) defines a completely-integrable system
for any real $k \neq 0$ which is also exactly-solvable.

After a gauge rotation $x_k\ =\ \Psi_0^{-1} ({\cal X}_k - c_k) \Psi_0$ this integral takes the algebraic form
\begin{equation}
\label{X_k-alg}
    {x}_k = -4k^2 u (t^k - u) \pa_u^2 - 4k^2 [(b+\frac{1}{2})t^k - (a+b+1)u]\pa_u \ ,
\end{equation}
where $c_k=k^2 (a+b)^2$ is the lowest eigenvalue of the integral ${\cal X}_k$.
It can be easily checked that ${x}_k$ has infinitely-many finite-dimensional invariant subspaces: it preserves the flag ${\cal P}^{(s)}$ for any $s \geq k$. The integral ${x}_k$  can be rewritten in the generators (\ref{gr}) as
\begin{equation}
\label{X_k-Lie}
    x_k = -4k J^3 R_k + 4 J^3 J^3 - 4k^2 (b+\frac{1}{2}) R_k +4k (a+b) J^3
    \ .
\end{equation}
The presence of the generator $R_k$ excludes the algebra (\ref{gr})-(\ref{grT}) for $s=k-1$ as hidden algebra (see the discussion above). It indicates that the hidden algebra of (\ref{X_k}) is $g^{(s)}$ for $s \geq k$. Hence, the hidden algebra of the quantum system (\ref{H}) with the integral (\ref{X_k}) is $g^{(s)}$ with $s \geq k$.

\section{\bf Superintegrability.}

The next question is the existence of an additional integral of motion
${\cal Y}_{2k}$ (presumably of order $2k$) for all integer values of $k$. If such an integral exists then the system (\ref{H}) is (maximally) superintegrable.
For $k=1$, the Smorodinsky-Winternitz system this integral ${\cal Y}_{2}$ was found long ago (see \cite{Fris:1965, Wint:1966} and \cite{TTW:2001}). It turned
to be a second order differential operator.
For $k=2$ and $\om=0$, which is the case of the so-called singular rational $BC_2$ model, the integral ${\cal Y}_{4}$ was found by Olshanetsky-Perelomov in the representation theory approach \cite{Olshanetsky:1983}. This integral is a  fourth order differential operator. For $k=3$ and $\om=\al=0$ (the so-called singular Calogero model) the corresponding integral is a third order differential operator \cite{Olshanetsky:1983}. For $\al \neq 0$ (the so-called singular Wolfes model) it
was mentioned in \cite{Olshanetsky:1983} that it has to be of the sixth order. For the general Wolfes model $\om\neq 0$ (the rational $G_2$ model in the Hamiltonian reduction method nomenclature) C.Quesne \cite{quesne} found this integral explicitly in the Dunkl operator formalism. The integral ${\cal Y}_{6}$ is a sixth order differential operator.

If the integral ${\cal Y}_{2k}$ exists it should have the same eigenfunctions as $H_k$. Hence by a gauge rotation and change of variables (15) we can obtain an operator ${y}_{2k}$
\begin{equation}
\label{Y_2k-gauge}
 y_{2k}=\Psi_0^{-1} ({\cal Y}_{2k}-C_{2k}) \Psi_0|_{t,u}\ ,
\end{equation}
such that ${y}_{2k}$ is a differential operator of some order in $t$ and $u$ with polynomial coefficients; $C_{2k}$ is the lowest eigenvalue of the integral ${\cal Y}_{2k}$. The described algebraic form ${y}_{2k}$ would be a consequence of the fact that  both $h_k$ (\ref{Lie}) and $y_{2k}$ should preserve the same flag of polynomials.

For the case $k=1$ the integral ${\cal Y}_2$ was found in \cite{Wint:1966}. In Cartesian coordinates ${\cal Y}_2$ is of  2nd order and it can be written as
\begin{equation}
\label{Int_1}
  {\cal Y}_2\ =\ \pa_x^2 - \om^2 x^2 - \frac{\al}{x^2} \ ,
\end{equation}

The algebraic form of the integral was calculated in \cite{TTW:2001}. In the coordinates (\ref{coord}) the integral (\ref{Int_1}) is
\begin{equation}
\label{Int_1_alg}
 \frac{y_2}{4}\ =\ (t-u) \pa^2_t + [\om(u-t)+a+\frac{1}{2}]\pa_t \ ,
\end{equation}
where the constant $C_2=-\om (2a+1)$.
The integral $y_2$ (\ref{Int_1_alg}) contains the term $u\pa_t$ which is present in the algebra $g^{(1)}$, see (\ref{grT}). It indicates unambiguously that for the case $k=1$ the hidden algebra should correspond to $s=1$. Hence, there is no ambiguity for the $k=1$ case. The hidden algebra is fixed and it is $g^{(1)} \equiv sl(3)$ which is generated by $gl(2)\ltimes R^2 \oplus T_1 \subset gl(3)$. The Lie algebraic form of $y_2$ is the following
\begin{equation}
\label{Int_1_Lie}
 \frac{y_2}{4}\ =\ J^2 J^1 - T_1 J^1 + \om T_1 - \om J^2 + (a+\frac{1}{2})J^1\ .
\end{equation}
We stress that the generator $T_1$ (see (\ref{grT})) appears explicitly in (\ref{Int_1_Lie}).

For the case $k=2$ (the rational $BC_2$ model) we find the higher integral ${\cal Y}_4$ explicitly,
\begin{align}
\label{Int_2}
{\cal Y}_4&=(\pa_x^2 - \om^2 x^2  - \pa_y^2 + \om^2 y^2 )^2+\bigg\{\pa_x^2,\frac{(x^2-y^2) \beta }{ x^2 y^2}-\frac{4(x^2+y^2) \al}{(x^2-y^2)^2}\bigg\}\non \\
+& \bigg\{\pa_x \pa_y,-\frac{16 x y \al}{(x^2-y^2)^2}\bigg\}
+\bigg\{\pa_y^2,-\frac{(x^2-y^2) \beta }{x^2 y^2}-\frac{4(x^2+y^2) \al }{(x^2-y^2)^2}\bigg\}\non \\
+ &\frac{16\al^2}{(x^2-y^2)^2}+\frac{(x^2-y^2)^2 \beta ^2}{x^4 y^4}+\frac{8\al  \beta }{x^2 y^2}-\frac{2 (x^4 +y^4) \beta  \om^2}{x^2y^2} \ ,
\end{align}
where $\{,\}$ denotes an anticommutator.
Making the gauge rotation $\Psi_0^{-1}({\cal Y}_4-C_4)\Psi_0$ and the change of variables (\ref{coord}), we arrive at the algebraic form of the integral
\begin{align*}
\label{Int_2_alg}
\frac{y_4}{16}&=(t^2-u)\pa_t^4-8(t^2-u)u\pa_t^2\pa_u^2+16(t^2-u)u^2\pa_u^4-2 [\om t^2
-(2 a +1)t-\om u ]\pa_t^3\\& - 4 [(2b+1)t^2- 2(a+b+1)u]\pa_t^2\pa_u+ 8 u [\om t^2
-(2 a +1)t-\om u ]\pa_t\pa_u^2\non \\
&+16 u [(2 b+3) t^2-2 (a+b+2) u]\pa_u^3+16 [\om^2t^2-3(2a +1)\om t - \om^2 u +
(2a +1)(2a+2 b+1)]\pa_t^2\non \\
&-4 [(2b+1)\om t^2-(2 a+1) (2 b+1) t- 2(a+b+1)\om u]\pa_t\pa_u\\
&+4 [(2b+1)(2b+3)t^2+(2 a+1)\om t u -2(2a^2+6ab+2b^2+8a+7b+5) u]\pa_u^2 \non
\end{align*}
\begin{equation}
\label{Int_2_alg}
 + \om (2 a+1)(\om t-2a-2b-1)\pa_t + 2 (2 a+1)(2b+1)(\om t-2a-2b-1)\pa_u \ ,
\end{equation}
where $C_4=4\om^2 [2a(a+1)-b(b-1)]$. The two terms $t^2 \pa_u$ and $u \pa^2_t$ in $y_4$ imply $s=2$. Hence, the hidden algebra for $k=2$ is $g^{(2)}$. The Lie-algebraic form of $y_4$ is the following
\begin{gather}
\label{Int_2_Lie}
\frac{y_4}{16}=J^2J^2J^1J^1-J^1J^1T_2+2J^1J^1J^3J^3+4J^3J^3R_2R_0
-4J^2J^2J^3R_0-2J^3J^3J^3R_0\non \\
-2\om J^2J^2J^1-2\om J^3J^3J^1 +2(2a+1)J^2J^1J^1-4(2b+1)J^2J^2R_0\non \\
+4\om J^3R_2J^1-4(2a+1)J^3R_1J^1+8(2b+3)J^3R_1R_1-8(a+b+2)J^3J^3R_0\non \\
+(2a+1)(2a+2b+1)J^1J^1-3\om (2a+1)J^2J^1+\om^2 J^2J^2+4\om (a+b+1)J^3J^1\non \\
 -4\om(2b+1)R_2J^1+64(2a+1)(2b+1)J^2R_0+2\om (2a+1)J^3R_1\non \\
 - 4(2a^2+6ab+2b^2+8a+7b+5)J^3R_0\non \\
+4(2b+1)(2b+3)R_2R_0 +2\om T_2J^1+8(a+b+1)T_2R_0\non \\
-(2a+1)(2a+2b+1)J^1+\om^2(2a+1)J^2\non \\
+2\om(2a+1)(2b+1)R_1-2(2a+1)(2b+1)(2a+2b+1)R_0-\om^2T_2 \ .
\end{gather}
The generator $T_2$ (see (\ref{grT})) again appears explicitly in (32).

For the case $k=3$ (the rational $G_2$ model) we find the higher integral ${\cal Y}_6$ explicitly by a straightforward (brute force) calculation. It is of 6th order (see App.A). Its lowest eigenvalue is
\begin{equation}
\label{C6}
  C_6\ =\ 4\om^3 (3a+3b+1)(5a^2+36ab-27b^2+a+45b+4)\ .
\end{equation}
Making the gauge rotation $\Psi_0^{-1}({\cal Y}_6-C_6)\Psi_0$ and the change of variables (\ref{coord}) we arrive at the algebraic form of the integral $y_6$ (see App.A).
The two elements $R_3=t^3 \pa_u$ and $T_3=u \pa^3_t$ are present in $y_6$ and unambiguously point to  $s=3$. Hence, the hidden algebra of the model at $k=3$ is $g^{(3)}$.

For the case $k=4$ we again find    the higher integral ${\cal Y}_8$ explicitly by a brute force calculation as an eigth order
differential operator (see App.B). Making the gauge rotation $\Psi_0^{-1}({\cal Y}_8-C_8)\Psi_0$ and the change of variables with
\[
C_8\ =\ 4\om^4 \big[ 3200 a^4 + 512 a^3 (31b+10) + 16 a^2 (206b+159)(2b-3)
\]
\begin{equation}
\label{C8}
 +16 a (310 b^3-187 b^2-443 b+105)+1133 b^4+150 b^3-176 b^2+493 b+4 \big]   \ ,
\end{equation}
we arrive at the algebraic form of the integral, $y_8$ (see App.B). The elements $R_4=t^4 \pa_u$ and $T_4=u \pa^4_t$ in $y_8$ imply $s=4$. The hidden algebra of the model for $k=4$ is $g^{(4)}$ which contains the generator $T_4$.

We were unable to prove the existence of the higher order integrals ${\cal Y}_{2k}$ for integer $k$ with $k>4$  due to the fast growing complexity of the brute force calculations. However, we feel justified in formulating the following conjecture.

{\bf
 Conjecture.
 \it An integral of motion ${\cal Y}_{2k}$ of the order $2k$ exists for the Hamiltonian (1) for all positive integer values of $k$.  In Cartesian coordinates ${\cal Y}_{2k}$ is a differential operator of the order $2k$ with rational coefficients. The gauge transformation (26) together with the change of variables (15) transforms ${\cal Y}_{2k}$ into the algebraic operator ${y}_{2k}$ that has polynomial coefficients. The integral ${y}_{2k}$ is an element of the order $2k$ in the enveloping algebra of the hidden algebra $g^{(k)}$. In particular, ${y}_{2k}$ contains the terms $4^k [(J^1)^k-T_k](J^1)^k$ which fix $k=s $ in the hidden algebra (19), (20).
 In the limit $\om = \al = 0$, the operator ${\cal Y}_{2k}(0,0,\beta)$ is reduced to the square of an operator of order $k$.
}

Our conjecture is based on the fact that the gauge rotated Hamiltonian $h_k$ (5) preserves the flag of polynomials (17), as do all the elements of the underlying hidden algebra $g^{(k)}$ (19), (20). All aspects of this Conjecture have been confirmed for $k=1,2,3$ and 4 for general $\om, \alpha, \beta$ as well as for $k=1,\ldots,6,8$ for $\om = \al = 0, \beta \neq 0$. The consideration of $k > 4$ for general $\om, \alpha, \beta$ requires a different approach, other than the brute force one. A proof of the conjecture could be based on a direct analysis of the commutation relations of the hidden algebra (19)-(20).

Any operator preserving this flag must lie in the enveloping algebra of $g^{(k)}$ for given $k$. The gauge rotated integrals $y_{2k}$ must hence have an algebraic form for all $k$, as exemplified by $k=1,2,3$ and 4.

The form of the  integrals ${\cal Y}_{2k}$ is not unique since we can modify it by adding  polynomials in the Hamiltonian (\ref{H}) and integral ${\cal X}_k$ (\ref{X_k}). Our convention is to require that the highest order terms in ${\cal Y}_{2k}$ should have the form
\begin{equation}
\big[ \text{Re}(\pa_1+\text{i}\pa_2)^k\big]^2\ .
 \end{equation}

The lower order terms in ${\cal Y}_{2k}$ could be further simplified by linear combinations with lower order polynomials in (\ref{H}) and (\ref{X_k}).  We also require that ${\cal Y}_{2k}$ be a hermitian operator and this implies that it will contain only even powers of the derivatives ($\pa_1^m\pa_2^n$, $m+n=0,2,4,\ldots ,2k$).

\section{\bf A Quasi-exactly solvable extension.}

Some years ago a new class of the Schroedinger equations was discovered  for which a finite number of eigenstates can be calculated by purely algebraic means. They were called {\it quasi-exactly-solvable} \cite{Turbiner:1987,Turbiner:1988}. These problems occupy an intermediate place between exactly-solvable problems and non-solvable ones. A large body of articles dedicated to these problems was published during the last 20 years. The articles have ranged from various branches of physics to pure mathematics.

Surprisingly, there exists a quasi-exactly solvable generalization of the Hamiltonian (\ref{H})
\[
 H_{k,{\cal N}}^{(qes)} (r,\varphi;\om, \al, \beta)\ =\ -\pa_r^2 - \frac{1}{r}\pa_r - \frac{1}{r^2}\pa_{\varphi}^2  + \la^2 r^6 + 2 \la \om r^4 + [\om^2-2\la (2{\cal N}+2+k(a+b))] r^2
\]
\begin{equation}
\label{H_qes}
 + \frac{\al k^2}{r^2 \cos^2 {k \varphi}} + \frac{\beta k^2}{r^2 \sin^2 {k \varphi}}\ ,
\end{equation}
(cf. \cite{Turbiner:1987,Turbiner:1988,Turbiner:2004}), where $\mbox{dim} {\cal P}_{{\cal N}}^{(k)} \approx \frac{{\cal N}^2}{2k}+1$ eigenstates can be found explicitly (algebraically). These algebraic eigenfunctions have the form of a polynomial $p(t,u)$ from the space ${\cal P}_{\cal N}^{(k)}$ (\ref{space_r}) multiplied by a factor $\Psi_0^{(qes)}$:
\begin{equation}
\label{psi0_qes}
   \Psi_0^{(qes)} \ =\ r^{(a+b)k}\ \cos^a {k \varphi}\ \sin^b {k \varphi}\ e^{-\frac{\om r^2}{2}- \frac{\la r^4}{4}}\ ,
\end{equation}
namely,
\begin{equation}
\label{psi_qes}
     \Psi_{alg}^{(qes)} \ =\ p(t,u) \Psi_0^{(qes)}  \ .
\end{equation}
Hence, the number of algebraic states is equal to the dimension of the space
${\cal P}_{\cal N}^{(k)}$.

The gauge-rotated Hamiltonian (\ref{H_qes}),
\[
 h_{k,{\cal N}}^{(qes)}\ =\ -(\Psi_0^{(qes)})^{-1}
 (H_{k,{\cal N}}^{(qes)}-E_0)\Psi_0^{(qes)}\ ,
\]
where $E_0$ is some parameter, in the variables (\ref{coord}) has the algebraic form:
\[
  h_{k,{\cal N}}^{(qes)}\ =   4 t \pa^2_t + 8k u \pa^2_{tu} + 4k^2 t^{k-1} u \pa^2_u
\]
\begin{equation}
\label{Lie_alg}
    + 4[\la t^2 - \om t + (a+b)k +1]\pa_t + [4 \la k t u -4\om k u + 2 k^2(2 b + 1) t^{k-1}]\pa_u - 4 \la {\cal N} t \ .
\end{equation}
It is easy to check that (\ref{Lie_alg}) preserves the space ${\cal P}_{\cal N}^{(k)}$ (\ref{space_r}). Hence, it can be rewritten in generators of the algebra (\ref{gr}), $gl(2)\ltimes R^{k+1}$ \cite{Turbiner:1994} and indeed we have
\[
  h_{k,{\cal N}}^{(qes)}/4 = (J^2_{\cal N} + 2J^3_{\cal N}) J^1 + k J^3_{\cal N} R_{k-1}
\]
\begin{equation}
\label{Lie_qes}
  + [(a+b)k+1+{\cal N}] J^1 -\om (J^2_{\cal N} + J^3_{\cal N}) - \la J^4_{\cal N} +
  \frac{k}{6} [2{\cal N} + 3k(2b+1)] R_{k-1}\ .
\end{equation}

Evidently, the QES problem is completely-integrable: ${\cal X}_k$ (see (\ref{X_k})) commutes with (\ref{H_qes}). The algebraic form of ${\cal X}_k$
after a gauge rotation with (\ref{psi0_qes}) in variables $(t,u)$
remains the same (\ref{X_k-alg}). The Lie-algebraic form (\ref{X_k-Lie})
is slightly modified
\begin{equation}
    x_k = -4k J^3_{\cal N} R_k + 4 J^3_{\cal N} J^3_{\cal N} - 4k [k (b+\frac{1}{2})+\frac{\cal N}{3}] R_k +4[k (a+b)+\frac{2 \cal N}{3} ] J^3_{\cal N} + \frac{4 {\cal N}^2}{9}
    \ .
\end{equation}

The question of the existence of a second integral and thus of the superintegrability of the Hamiltonian (\ref{H_qes}) remains open.

\section{Conclusions.}

We have restricted to the case of a Schroedinger equation in a two dimensional Euclidean space $E_2$ and to the Hamiltonians allowing separation of variables in polar coordinates. The feature underlying the exact solvability, the complete integrability and the conjecture of maximal superintegrability is the existence of a hidden Lie algebra of differential operators. All elements of the hidden algebra and hence also of its enveloping algebra preserve an infinite flag of finite dimensional subspaces of the space of wave functions.

The Hamiltonians and the integrals of motion of the entire family (\ref{H}) considered in this article are also elements of the enveloping algebra of $g^{(k)}$. The family contains {\it all} currently known superintegrable systems in $E_2$ that are separable in polar coordinates. It would be important to clarify whether the Hamiltonian (1) can be obtained
by a Hamiltonian reduction procedure. This is the case for $k=1,2,3$.

The first problem that remains open is to prove our conjecture, namely that the Hamiltonian (\ref{H}) is superintegrable for all integers values of $k$. Another important question is that of the classical limit of the system with Hamiltonian (1). For $k=1,2$ and 3 these systems are all superintegrable. Chanu et al. \cite{chanu} have considered the classical case for $\om = \al = 0$ and $k=2n+1$ and have conjectured that it is superintegrable for all integer $n$. We think that the classical limit of
(1) is actually superintegrable for all values of $\om, \al$ and $k$.
We plan to verify this conjecture directly by calculating the trajectories for the classical systems. If the systems are (maximally) superintegrable then all bounded trajectories must be closed and the motion must be periodic \cite{nekhoro}.

The direct construction of the higher order integrals ${\cal Y}_{2k}$ for $k \geq 5$ seems intractable. More promising approaches would either involve an efficient use of the hidden algebra $g^{(k)}$ or possibly the use of Dunkl operators \cite{dunkl} as suggested for the Calogero model in \cite{poly} and for the Wolfes model in \cite{quesne}.

The close relation between exact solvability and maximal superintegrability has also been exemplified in $n$ dimensions \cite{rod,temp,evans}. A very complete review of quantum completely-integrable systems in $n$ dimensions was recently given by Oshima \cite{osh}.
For some cases these systems are known to be exactly-solvable. It would be
of great interest to investigate their possible (maximal) superintegrability.

\begin{acknowledgments}
A.V.T. wants to express his deep gratitude to S.P.Novikov for useful discussions.
A.V.T. thanks the CRM, University of Montreal, Canada where this work was started and IHES, Bures-sur-Yvette, France where it was completed for their kind hospitality extended to him. The research of A.V.T. is supported in part by DGAPA grant IN121106 (Mexico). A.V.T. thanks the University Program FENOMEC (UNAM, Mexico) for a partial support. The research of P.W. was partially supported by a research grant from NSERC of Canada.
\end{acknowledgments}

\appendix\section{$k=3$ case}

The integral for $k=3$ is
\begin{align*}
\begin{split}
&\mathcal{Y}_6=( \pa_x^2-\om^2x^2)^3-6( \pa_x^2-\om^2x^2)^2(\pa_y^2-\om^2y^2)+9( \pa_x^2-\om^2x^2)(\pa_y^2-\om^2y^2)^2
\\
&+\bigg\{\partial_x^4,-\frac{9(-3 x^4+6 y^2 x^2+y^4) \beta }{y^2(y^2-3 x^2 )^2}-\frac{27 (x^2+y^2)^2 \alpha }{2x^2 (x^2-3  y^2)^2}\bigg\}+\bigg\{\partial_x^3\partial_y-\frac{72 x y \beta }{(y^2-3 x^2)^2}-\frac{216 x y \alpha }{(x^2-3 y^2)^2}\bigg\}
\\
&+\bigg\{\pa_x^2\pa_y^2,\frac{27 (-3 x^4+6  x^2y^2+y^4) \beta }{(y^3-3 x^2
  y)^2}-\frac{54 (x^4+4x^2y^2-y^4) \al}{(x^3-3 x y^2)^2}+6y^2 \om^2\bigg\}\\
&+\bigg\{\pa_x\pa_y^3,\frac{216 x y \beta }{(y^2-3 x^2)^2}+\frac{72 x y \alpha }{(x^2-3 y^2)^2}-12xy\omega^2\bigg\}+\bigg\{\partial_y^4,-\frac{9\alpha }{2 x^2}+6x^2\omega^2\bigg\}\\
&+\bigg\{\partial_x^2,\frac{18 (3 x^8+44 y^2 x^6+42 y^4 x^4-36 y^6 x^2+27 y^8) \alpha
   }{x^4 (x^2-3 y^2)^4}+\frac{81(3 x^4-6 y^2 x^2-y^4)^2 \beta ^2}{2 y^4 (3
   x^2-y^2)^4}\\
    &+\frac{81(3 x^8-52 y^2 x^6+18 y^4 x^4+12 y^6 x^2+3 y^8) \alpha ^2}{2
   x^4 (x^2-3 y^2)^4}+\frac{2 (2 x^6+15 y^2 x^4+18 y^4 x^2-27 y^6) \alpha  \omega^2}{
   x^2 (x^2-3 y^2)^2}\\
   &+\frac{162 (x-y) (x+y) (3 x^6-19 y^2 x^4-7 y^4 x^2-y^6) \alpha  \beta
   }{x^2 y^2 (x^2-3 y^2)^2 (3 x^2-y^2)^2} -\frac{54 x^2 (x-y) (x+y) (x^2-7 y^2) \beta  \omega^2}{y^2 (3 x^2-y^2)^2}\\
   &-6y^4\omega^4\bigg\}\\
    &+\bigg\{\partial_x\partial_y,-\frac{1152 x y (x^2+3 y^2) \alpha }{(x^2-3 y^2)^4}-\frac{648 x (3 x^4-6 y^2 x^2-y^4) \beta ^2}{y (y^2-3
   x^2)^4}-\frac{648 y (x^4+6 y^2 x^2-3 y^4) \alpha ^2}{x (x^2-3
   y^2)^4}\\
    &-\frac{648 (x-y) (x+y) (x^4+10 y^2 x^2+y^4) \alpha  \beta }{x y
   (3 x^4-10 y^2 x^2+3 y^4)^2}+\frac{108 x(x-y) (x+y) (x^2+y^2) \beta  \omega^2}{y (y^2-3 x^2)^2}\\
   &+\frac{216 x y (x^2+y^2) \alpha  \omega^2}{(x^2-3y^2)^2}+12 x y^3 \omega^4\bigg\}\\
   &+\bigg\{\partial_y^2,\frac{81 (x^8-6 y^4 x^4+24 y^6 x^2-3 y^8) \alpha ^2}{x^4 (x^2-3 y^2)^4}+\frac{144 (x^4+18 y^2 x^2+9 y^4) \alpha
   }{(x^2-3 y^2)^4}+\frac{2592 x^2 y^2 \beta ^2}{(3 x^2-y^2)^4}\\
   &-\frac{(23 x^6+159 y^2 x^4+45 y^4 x^2-27 y^6) \alpha  \omega^2}{x^2 (x^2-3 y^2)^2}+\frac{81 (x^8-6 y^4 x^4+56 y^6 x^2-3
   y^8) \alpha  \beta }{x^2 y^2 (x^2-3 y^2)^2 (3 x^2-y^2)^2}
   \\
   &-\frac{27 (x^6-11 y^2 x^4+19 y^4 x^2-y^6) \beta \om^2}{y^2 (3 x^2-y^2)^2}-6 x^2y^2 \om^4+2\om^2\bigg\}\\
   &-\frac{180 (43 x^{12}+1914 y^2 x^{10}+5805 y^4 x^8+972 y^6 x^6+405 y^8 x^4-486 y^{10} x^2+243 y^{12}) \al}{x^6 (x^2-3 y^2)^6}\\
   &+\frac{1296 \beta ^2}{(y^3-3 x^2 y)^2}+\frac{1296 (x^6+21 y^2 x^4-9 y^4 x^2+3 y^6) \al \beta }{x^2 y^2 (x^2-3 y^2)^2 (3 x^2-y^2)^2}-\frac{729\al^3 (x^2+y^2)^6}{x^6 (x^2-3 y^2)^6}\\
   &-\frac{324 (5 x^{12}+30 y^2
   x^{10}+399 y^4 x^8-332 y^6 x^6+291 y^8 x^4-18 y^{10} x^2+9 y^{12}) \al^2}{x^6 (x^2-3 y^2)^6}\\
   &-\frac{1458 \al^2 \beta (x^2+y^2)^6}{x^4 y^2 (x^2-3 y^2)^4 (3 x^2-y^2)^2}-
   \frac{729\al \beta^2  (x^2+y^2)^6}{x^2 y^4 (x^2-3 y^2)^2 (3 x^2-y^2)^4} +\frac{72 (x^2+y^2)^2 \al \om^2}{(x^3-3 x y^2)^2}\\
\end{split}
\end{align*}

\begin{align}
\begin{split}
   &+\frac{243(x^2+y^2)^2 (x^6-24 y^2 x^4+21 y^4 x^2-2 y^6) \beta ^2 \omega^2}{9 x^4 (x^2-3 y^2)^4}+\frac{414 (x^2+y^2)^2 \alpha  \beta  \omega^2}{ y^2 (3 x^2-y^2)^2}\\
   &+\frac{9 (x^2+y^2)^2 (19 x^6+372 y^2 x^4-153 y^4 x^2+54 y^6) \alpha ^2 \omega ^2}{x^4 (x^2-3 y^2)^4}+\frac{180 (x^2+y^2)^2 \beta  \omega^2}{y^2 (3 x^2-y^2)^2}\\
   &+\frac{54 (x-y) (x+y) (x^2+y^2) (x^2-4 y x+y^2) (x^2+4 y x+y^2) \beta  \omega^4}{y^2 (3
   x^2-y^2)^2}\\
   &+\frac{(x^2+y^2) (19 x^6+129 y^2 x^4+9 y^4 x^2+27 y^6) \al \om^4}{ x^2 (x^2-3 y^2)^2}-4y^2\om^4 \ .
   \end{split}
\end{align}
The algebraic form $y_6=\Psi_0^{-1}(\mathcal{Y}_6-C_6)\Psi_0$ after the change of variables $t=r^2$ and $u=r^6\sin^23\varphi$ is
\begin{align*}
y_6
&=64(t^3-u)\pa_t^6 -3456  (t^3-u)t u \pa_t^4\pa_u^2 -6912 (t^3-u) u^2\pa_t^3\pa_u^3+ 46656 (t^3-u)  t^2u^2\pa_t^2\pa_u^4
\\
&+186624 (t^3-u) tu^3\pa_t\pa_u^5+ 186624 (t^3-u) u^4\pa_u^6
\\
&-96 [2\om t^3-3(2a+1) t^2-2\om u]\pa_t^5-1728[(2b+1)t^3-2(a+b+1)u]\,t\,\pa_t^4\pa_u
\\
&+1728[4\om t^4-(12a+6b+19)t^3-4\om t u+2(3a+3b+8)u]\,u\,\pa_t^3\pa_u^2
\\
&+5184 \big[9(2 b+3) t^5+2 \om t^3 u-3 (8 a+6 b+13) t^2u-2 \om u^2 \big]\,u\,\pa_t^2\pa_u^3
\\
&-23328 [2\om t^4-(6a+20b+57) t^3-2\om t u+4(5 a+5 b+16)u]\,t u^2\,\pa_t\pa_u^4
\\
&-46656 [2\om t^4-(6a+12b+47) t^3-2 \om t u +2(6a+6b+25)u]\,u^3\,\pa_u^5
\\
&+48\big[4\om^2 t^3-15(2a +1)\om t^2+6(2a+1)(3a+3b+2)t -4\om^2 u\big]\partial_t^4+864 \big[4(2b+1)\om t^4\\
&-(2b+1)(12a+2b+13) t^3-8 (a+b+1)\om t u+2(4a^2+3(2b+3)a+2b^2+11b+7) u\big]\pa_t^3\pa_u
\\
&+432 \big[27(2b+3)(2b+1) t^5-8\om^2 t^4 u+ 6(12 a+6 b+19)\om t^3 u
\\
&-9 (16 a^2+16 (3 b+5) a+12 b^2+48 b+45)t^2 u+8 \om^2 t u^2 -12 (3 a+3 b+8) \om u^2  \big]\pa_t^2\pa_u^2
\\
&-7776 \big[6(2b+3)\om t^4-3(2b+3)(6a+8b+27)t^3-2(8a+6b+13)\om t u
\\
&+2 (24 a^2+2 (30 b+71) a+24 b^2+138 b+173) u\big]t u\,\pa_t\pa_u^3\\
&+3888  \big[4\om^2 t^5-3(6a+20b+57)\om t^4+6(4a^2+(30b+83)a+26b^2+179b+288) t^3
\\
&-4 \om^2 t^2 u+12(5a+5b+16)\om t u-12(13 a^2+(30b+97)a+13b^2+97b+169)u \big]u^2\pa_u^4
\\
&-\big[64\om^3 t^3-576(2a+1)\om^2 t^2+576(6 a^2+6 b a+7 a+3 b+2)\om t-64\om^3 u\\
&-96(2a+1)(3a+3b+2)(3a+3b+1)\big]\pa_t^3
\\
&-\big[1728(2b+1)\om^2 t^4-1296(2b+1)(12a+2b+13)\om t^3+7776(2a+1)(2b+1)
(2a+2b+3)t^2
\\
&-3456 (a+b+1)\om^2 t u+2592(4a^2+3(2b+3)a+2b^2+11b+7)\om u\big]\pa_t^2\pa_u
\end{align*}
\begin{align}
\begin{split}
&-\big[11664 (2b+3)(2b+1)\om t^5-5832(6a+4b+17)(2b+3)(2b+1)t^4+1728(6a+5)\om^2 t^3 u
\\
& -3888 (16a^2+16(3b+5)a+12 b^2+48b+45)\om t^2 u+2592[36 a^3+6(36b+53)a^2
\\
&+(216b^2+792b+685)a+36b^3+306b^2+621b+362]t u-3456 \om^2 u^2 \big]\pa_t\pa_u^2
\\
&+\big[15552 (2 b+3)\om^2 t^5 u-11664(2b+3)(6a+8b+27)\om t^4 u+432 (648 b^3+5724 b^2
+15282 b
\\
&+216 a^2 (2 b+3)+108a(12b^2+56b+57)+12231) t^3 u -1728\om^3 t^3 u^2
-31104 (a+b+2) \om^2 t^2  u^2
\\
&+7776(24a^2+2(30b+71)a+24b^2+138b+173) \om t u^2-864(324a^3+108(13b+30)a^2
\\
&+6(234b^2+1242b+1597)a+324b^3+3240b^2+9558b+8679) u^2+1728\om^3 u^3 \big]\pa_u^3
\\
&-\big[144(2a+1)\om^3 t^2-8(247a^2+216 ba+293a+27 b^2+81b+80) \om^2 t
\\
&+144(2a+1)(3a+3b+2)(3a+3b+1) \om\big]\pa_t^2
\\
&-\big[864(6a+5)(2b+1)\om^2 t^3-7776(2a+1)(2b+1)(2a+2b+3) \om t^2 +2592(2a+1)(2b+1)
\\
&(3a+3b+2)(3a+3b+5)t -48(31a^2+41a+27b^2+45b+44)\om^2 u\big]\pa_t\pa_u
\\
&+\big[3888(2b+3)(2b+1) \om^2 t^5-2916(2b+3)(2b+1)(6a+4b+17) \om t^4
\\
&+5832(2b+3)(2b+1)(a+b+4)(4a+2b+7) t^3 -1296 (2b+3) \om^3 t^3 u
\\
&-72 (185 a^2+(648 b+1003) a+189 b^2+783 b+622)\om^2 t^2 u +1296(36a^3+6(36b+53)a^2
\\
&+(216b^2+792b+685)a+36b^3+306b^2+621b+362)\om t u+2592(a+b+2)\om^3 u^2
\\
&-1296 (36a^4+324a^3b+450a^3+576b^2a^2+1962a^2b+1619a^2+324b^3a
   +1962ab^2+3678ab
\\
&+2213a+36b^4+450b^3+1611b^2+2181b+992)u\big]\pa_u^2
\\
&-(31a^2+27b^2+41a-27b+8) \big[8\om t -8(3a+3b+1)\big] \om^2\pa_t
\\
&-12\big[18(2b+3)(2b+1) \om^3 t^3+3(2b+1)(185a^2-27b^2+216ab+135b+355a+136) \om^2 t^2
\\
&-108(2b+1)(2a+1)(3a+3b+5)(3a+3b+2) \om t
\\
&-2(41 a^2 +108 b a+9 b^2+121a+153 b+82) \om^3 u
\\
&+108(2a+1)(2b+1)(a+b+2)(3a+3b+1)(3a+3b+2)\big]\pa_u \ ,
\end{split}
\end{align}
where $C_6$ is the lowest eigenvalue of $\mathcal{Y}_6$ (see (\ref{C6})).

\section{$k=4$ case}
The integral for $k=4$ is
\begin{align*}
\mathcal{Y}_8&=\Big((\pa_x^2-\om^2x^2)^2-6(\pa_x^2-\om^2x^2)(\pa_y^2-\om^2y^2)
+(\pa_y^2-\om^2y^2)^2\Big)^2
\\
&+\bigg\{\partial_x^6,-\frac{32(x^2+y^2)^3\al}{(x^4-6x^2y^2+y^4)^2}
+\frac{2 (3 x^6-11 y^2 x^4+y^4 x^2-y^6) \beta}{x^2y^2(x^2-y^2)^2}\bigg\} +\bigg\{\pa_x^5\pa_y,-\frac{64 x y \beta }{(x^2-y^2)^2}
\\
&-\frac{256xy(5x^4-2x^2y^2+y^4)\alpha}{(x^4-6x^2y^2+y^4)^2} \bigg\} +\bigg\{\pa_x^4\pa_y^2,-\frac{32(11x^6+81x^4y^2-63x^2y^4-5y^6)}{(x^4-6x^2y^2+y^4)^2}
\\
&-\frac{2 (19 x^6-67 y^2 x^4+17 y^4 x^2-9 y^6) \beta }{x^2 y^2(x^2-y^2)^2} +32\om^2y^2\bigg\}+\bigg\{\pa_x^3\pa_y^3,\frac{512xy(3x^4-2x^2y^2+3y^4)
\al}{(x^4-6x^2y^2+y^4)^2}
\\
&+\frac{384 x y \beta }{(x^2-y^2)^2}-64\om^2xy\bigg\} +\bigg\{\pa_x^2\pa_y^4,\frac{32(5x^6+63x^4y^2-81x^2y^4-11y^6)\al}{(x^4-6x^2y^2+y^4)^2} +32\om^2x^2
\\
&+\frac{2 (9 x^6-17 y^2 x^4+67 y^4 x^2-19 y^6) \beta }{x^2 y^2 (x^2-y^2)^2}\bigg\} +\bigg\{\pa_x\pa_y^5,-\frac{256xy(x^4-2x^2y^2+5y^4)\alpha}{(x^4-6x^2y^2+y^4)^2}
\\
&-\frac{64 x y \beta }{(x^2-y^2)^2}\bigg\}+\bigg\{\partial_y^6,-\frac{2 (x^6-y^2 x^4+11 y^4 x^2-3 y^6) \beta }{x^2 y^2 (x^2-y^2)^2} -\frac{32(x^2+y^2)^3\al}{(x^4-6x^2y^2+y^4)^2}\bigg\}
\\
&+\bigg\{\pa_x^4,\frac{384 (x^2+y^2)^2 (3 x^8+92 y^2 x^6-142 y^4 x^4+92 y^6 x^2
+3 y^8) \alpha }{(x^4-6 y^2 x^2+y^4)^4}
\\
&-\frac{24 (3 x^8+28 y^2 x^6-2 y^4 x^4+4 y^6 x^2-y^8) \beta }{(x^3-x y^2)^4}
+\frac{256 (x^8-10 y^2 x^6+37 y^4 x^4-16 y^6 x^2) \al \om^2}{(x^4-6 y^2 x^2+y^4)^2}
\\
 &+\frac{256 (3 x^{12}-182 y^2 x^{10}+205 y^4 x^8-52 y^6 x^6+77 y^8 x^4+10 y^{10} x^2
 +3 y^{12}) \al^2}{(x^4-6 y^2x^2+y^4)^4}
\\
 &+\frac{32 (11 x^{12}-158 y^2 x^{10}+221 y^4 x^8-68 y^6 x^6+53 y^8 x^4+2 y^{10} x^2+3 y^{12}) \alpha  \beta }{x^2 y^2 (x^6-7 y^2 x^4+7 y^4 x^2-y^6)^2}
\\
 &+\frac{(19 x^{12}-134 y^2 x^{10}+237 y^4 x^8-84 y^6 x^6+29 y^8 x^4-6 y^{10} x^2+3 y^{12}) \beta ^2}{x^4 y^4 (x^2-y^2)^4}
\\
  & -\frac{2 (9 x^8-50 y^2 x^6+98 y^4 x^4-18 y^6 x^2+9 y^8) \beta  \omega ^2}{x^2 y^2 (x^2-y^2)^2}\bigg\}
\\
  &+\bigg\{\pa_x^3\pa_y,\frac{6144 x (7 y^{11}+37 x^2 y^9-34 x^4 y^7+34 x^6 y^5-37 x^8 y^3-7 x^{10} y) \alpha }{(x^4-6 y^2 x^2+y^4)^4}-\frac{768 x y (x^2+y^2) \beta }{(x^2-y^2)^4}
\\
  &-\frac{8192 x (-y^{11}-5 x^2 y^9+10 x^4 y^7-30 x^6 y^5+23 x^8 y^3+3 x^{10} y) \al ^2}{(x^4-6 y^2 x^2+y^4)^4}+\frac{64 x y (x^2-3 y^2) \beta \om^2}{(x^2-y^2)^2}
\\
  &-\frac{128 (3 x^6-11 y^2 x^4+y^4 x^2-y^6) \beta ^2}{x y (x^2-y^2)^4} +\frac{512 x
  (- y^7- x^2 y^5+5 x^4 y^3+5 x^6 y) \al \om^2}{(x^4-6 y^2 x^2+y^4)^2}
\\
  &-\frac{512 (3 x^{10}+35 y^2 x^8-74 y^4 x^6+14 y^6 x^4-9 y^8 x^2-y^{10}) \al \beta}
  {x y (x^6-7 y^2 x^4+7 y^4 x^2-y^6)^2}-64xy^3\om^4\bigg\}
\end{align*}

\begin{align*}
   &+\bigg\{\pa_x^2\pa_y^2,\frac{768 (x^2+y^2)^2 (3 x^8+92 y^2 x^6-142 y^4 x^4+92 y^6 x^2+3 y^8) \alpha }{(x^4-6 y^2 x^2+y^4)^4}
\\
   &-\frac{72 (x^{12}-4 y^2 x^{10}-y^4 x^8-56 y^6 x^6-y^8 x^4-4 y^{10} x^2+y^{12}) \beta }{x^4 y^4 (x^2-y^2)^4}
\\
   &-\frac{512 (x^{12}+14 y^2 x^{10}+303 y^4 x^8-956 y^6 x^6+303 y^8 x^4+14 y^{10} x^2+y^{12}) \alpha ^2}{(x^4-6 y^2 x^2+y^4)^4}
\\
   &-\frac{2 (9 x^{12}-34 y^2 x^{10}+39 y^4 x^8-1372 y^6 x^6+39 y^8 x^4-34 y^{10} x^2
   +9 y^{12}) \beta ^2}{x^4 y^4 (x^2-y^2)^4}
\\
   &-\frac{64 (5 x^{12}-10 y^2 x^{10}+171 y^4 x^8-1164 y^6 x^6+171 y^8 x^4-10 y^{10} x^2
   +5 y^{12}) \al \beta }{x^2 y^2 (x^6-7 y^2 x^4+7 y^4 x^2-y^6)^2}
\\
   &+\frac{4 (3x^8-14 y^2 x^6-218 y^4 x^4-14 y^6 x^2+3 y^8) \beta \om^2}{x^2 y^2 (x^2-y^2)^2}
\\
 &+\frac{512  (x^8-15x^6y^2-32x^4y^4-15x^2y^2+y^8) \al \om^2}{(x^4-6 y^2 x^2+y^4)^2}+128x^2y^2\om^4+64\om^2 \bigg\}
\\
 &+\bigg\{\pa_x\pa_y^3,\frac{6144 (7 y x^{11}+37 y^3 x^9-34 y^5 x^7+34 y^7 x^5-37 y^9 x^3 -7 y^{11} x) \alpha }{(x^4-6 y^2 x^2+y^4)^4}
\\
&-\frac{768 x y (x^2+y^2) \beta }{(x^2-y^2)^4}+\frac{8192 x y (x^{10}+5 y^2 x^8-10 y^4 x^6+30 y^6 x^4-23 y^8 x^2-3 y^{10}) \al^2}{(x^4-6 y^2 x^2+y^4)^4}
\\
&+\frac{128 (x^6-y^2 x^4+11 y^4 x^2-3 y^6) \beta ^2}{x y (x^2-y^2)^4}
-\frac{512 xy(x^6+x^4y^2-5x^2y^4-5y^6) \al \om^2}{(x^4-6 y^2 x^2+y^4)^2}
\\
&+\frac{512 (x^{10}+9 y^2 x^8-14 y^4 x^6+74 y^6 x^4-35 y^8 x^2-3 y^{10}) \al \beta}
{x y (x^6-7 y^2 x^4+7 y^4 x^2-y^6)^2}-\frac{64y (x^2 + y^2) \beta \om ^2}{x(x^2-y^2)}-64x^3y\om^4 \bigg\}
\\
&+\bigg\{\pa_y^4,\frac{384 (x^2+y^2)^2 (3 x^8+92 y^2 x^6-142 y^4 x^4+92 y^6 x^2+3 y^8) \alpha }{(x^4-6 y^2 x^2+y^4)^4}
\\
&+\frac{24 (x^8-4 y^2 x^6+2 y^4 x^4-28 y^6 x^2-3 y^8) \beta }{(y^3-x^2 y)^4}
-\frac{256 y^2 (16x^6-37 y^2 x^4+10y^4 x^2- y^6) \al \om^2}{(x^4-6 y^2 x^2+y^4)^2}
\\
&+\frac{256 (3 x^{12}+10 y^2 x^{10}+77 y^4 x^8-52 y^6 x^6+205 y^8 x^4-182 y^{10} x^2+3 y^{12}) \al^2}{(x^4-6 y^2 x^2+y^4)^4}
\\
&+\frac{32 (3 x^{12}+2 y^2 x^{10}+53 y^4 x^8-68 y^6 x^6+221 y^8 x^4-158 y^{10} x^2+11 y^{12}) \al \beta }{x^2 y^2 (x^6-7 y^2x^4+7 y^4 x^2-y^6)^2}
\\
&+\frac{(3 x^{12}-6 y^2 x^{10}+29 y^4 x^8-84 y^6 x^6+237 y^8 x^4-134 y^{10} x^2
+19 y^{12}) \beta^2}{x^4 y^4 (x^2-y^2)^4}
\\
&-\frac{2 (9 x^8-18 y^2 x^6+98 y^4 x^4-50 y^6 x^2+9 y^8) \beta \om^2}{x^2 y^2 (x^2-y^2)^2}\bigg\}
\\
   &+\bigg\{\pa_x^2,-\frac{15360 (17 x^{18}+1467 y^2 x^{16}+7140 y^4 x^{14}-4932 y^6 x^{12}+318 y^8 x^{10}+666 y^{10} x^8)\alpha }{(x^4-6 y^2 x^2+y^4)^6}
\\
   &-\frac{15360(-1068 y^{12} x^6+1260 y^{14} x^4+249y^{16} x^2+3 y^{18}) \al}
   {(x^4-6 y^2 x^2+y^4)^6}\\&  - \frac{120(165 y^8 x^{10}+333 y^{10} x^8
   -87 y^{12} x^6+75 y^{14} x^4-30 y^{16} x^2+5 y^{18}) \beta}{x^6 y^6 (x^2-y^2)^6}
\end{align*}

\begin{align*}
   &+\frac{120 (3 x^{18}-18 y^2 x^{16}+45 y^4 x^{14}-81 y^6 x^{12})\beta }{x^6 y^6 (x^2-y^2)^6}-\frac{2048 x^2 (31 x^{16}+264 y^2 x^{14}-7380 y^4 x^{12})\alpha^2}{(x^4-6 y^2 x^2+y^4)^6}\\
&-\frac{2048 x^2(-17256 y^6 x^{10}+26538 y^8 x^8-16392 y^{10} x^6+8628 y^{12} x^4-1560 y^{14} x^2+303
   y^{16}) \al^2}{(x^4-6 y^2 x^2+y^4)^6}\\
   &-\frac{128 (33 x^{20}-714 y^2 x^{18}+7551 y^4 x^{16}-17856 y^6 x^{14}+21866 y^8 x^{12}-19604 y^{10} x^{10}+7470 y^{12} x^8) \alpha  \beta }{y^2 (x^3-x y^2)^4 (x^4-6 y^2 x^2+y^4)^3}
\\
   &-\frac{128(-2160 y^{14} x^6+453y^{16} x^4-114 y^{18} x^2+3 y^{20})\al\beta}
   {y^2 (x^3-x y^2)^4 (x^4-6 y^2 x^2+y^4)^3}+\frac{8 (3 x^{18}-18 y^2 x^{16}+354 y^4 x^{14}) \beta ^2}{x^6 y^6 (x^2-y^2)^6}
\\
   &+\frac{8(-1281 y^6 x^{12}+1242 y^8 x^{10}-1551 y^{10} x^8+282 y^{12} x^6-87 y^{14} x^4+39 y^{16} x^2-7 y^{18})\beta^2}{x^6 y^6 (x^2-y^2)^6}
\\
   &-\frac{8192 (x^2+y^2)^4 (x^{10}+45 y^2 x^8-46 y^4 x^6+34 y^6 x^4-3 y^8 x^2+y^{10}) \alpha ^3}{(x^4-6 y^2x^2+y^4)^6}
\\
  &+\frac{2 (x^2+y^2)^4 (3 x^{10}-29 y^2 x^8+70 y^4 x^6-18 y^6 x^4+7 y^8 x^2-y^{10})
  \beta^3}{x^6 y^6 (x^2-y^2)^6}
\\
   &+\frac{512 (x^2+y^2)^4 (x^{10}-119 y^2 x^8+162 y^4 x^6-86 y^6 x^4+13 y^8 x^2
   -3 y^{10}) \al^2 \beta }{x^2 y^2 (x^2-y^2)^2 (x^4-6 y^2 x^2+y^4)^4}
\\
   &+\frac{32 (x^2+y^2)^4 (5 x^{10}-103 y^2 x^8+186 y^4 x^6-70 y^6 x^4+17 y^8 x^2-3 y^{10}) \alpha  \beta^2}{x^4 y^4(x^2-y^2)^4 (x^4-6 y^2 x^2+y^4)^2}
   \\
 &-\frac{256 (4 x^{14}+9 y^2 x^{12}-1030 y^4 x^{10}+775 y^6 x^8-512 y^8 x^6+167 y^{10} x^4-3 y^{14}) \al \om^2}{(x^4-6 y^2 x^2+y^4)^4}
\\
    &+\frac{8 (13 x^{12}+16 y^2 x^{10}+367 y^4 x^8+112 y^6 x^6+91 y^8 x^4-32 y^{10} x^2
    +9 y^{12}) \beta \om^2}{x^4y^2 (x^2-y^2)^4}
    \\
 &-\frac{512 (x^2+y^2)^2 (13 x^{10}-315 y^2 x^8+506 y^4 x^6+354 y^6 x^4-215 y^8 x^2+9 y^{10}) \al^2 \om^2}{(x^4-6 y^2 x^2+y^4)^4}
\\
  &-\frac{2 (x^2+y^2)^2 (3 x^{10}-25 y^2 x^8+26 y^4 x^6-142 y^6 x^4+51 y^8 x^2-9 y^{10}) \beta ^2 \om^2}{x^4 y^4(x^2-y^2)^4}
\\
   &-\frac{128 (x^2+y^2)^2 (37 x^8-25 y^2 x^6+55 y^4 x^4-7 y^6 x^2+4 y^8) \al \beta  \om^2}{(x^7-7 y^2 x^5+7 y^4x^3-y^6 x)^2}
\\
   &-96 \om^4 x^2 -\frac{32 (13 x^{10}-156 y^2 x^8+630 y^4 x^6+76 y^6 x^4+5 y^8 x^2+24 y^{10}) \al \om^4}{(x^4-6 y^2 x^2+y^4)^2}
\\
&+\frac{2 (9 x^{10}-99 y^2 x^8+171 y^4 x^6-212 y^6 x^4-10 y^8 x^2-19 y^{10}) \beta
 \om^4}{x^2 y^2 (x^2-y^2)^2}+32 x^2 y^4 \om^6\bigg\}
\\
&+\bigg\{\pa_x\pa_y,-\frac{61440 x (51 y^{17}+936 x^2 y^{15}+756 x^4 y^{13}
 -1512 x^6 y^{11}+1330 x^8 y^9) \alpha }{(x^4-6 y^2 x^2+y^4)^6}\\
&-\frac{61440 x(-1512 x^{10} y^7+756 x^{12} y^5+936 x^{14} y^3+51 x^{16}
   y)\alpha}{(x^4-6 y^2 x^2+y^4)^6}+\frac{30720 x y (3 x^4+10 y^2 x^2+3 y^4)
   \beta }{(x^2-y^2)^6}
\\
   &-\frac{4096 x (39 y^{17}+888 x^2 y^{15}+3396 x^4 y^{13}-18360 x^6 y^{11}
   +36010 x^8 y^9-18360 x^{10} y^7) \alpha ^2}{(x^4-6 y^2 x^2+y^4)^6}
\end{align*}

\begin{align*}
   &-\frac{4096 x(3396 x^{12} y^5+888 x^{14} y^3+39 x^{16} y)\alpha^2}{(x^4-6 y^2 x^2+y^4)^6}+\frac{16 (9 x^{12}+258 y^2 x^{10}-585 y^4 x^8+3196 y^6 x^6) \beta ^2}{x^3 y^3
   (x^2-y^2)^6}\\
   &+\frac{16(-585 y^8 x^4+258 y^{10} x^2+9 y^{12})\beta^2}{x^3 y^3
   (x^2-y^2)^6}-\frac{65536 x y (x^2-y^2)^2 (x^2+y^2)^4 (x^4+10 y^2 x^2+y^4) \alpha ^3}{(x^4-6 y^2 x^2+y^4)^6}\\
   &+\frac{256 (7 x^{20}+104 y^2 x^{18}+243 y^4 x^{16}-3792 y^6 x^{14}+18438 y^8 x^{12}-33072 y^{10} x^{10}) \alpha  \beta }{x^3 y^3 (x^2-y^2)^4 (x^4-6 y^2 x^2+y^4)^3}\\
   &+\frac{256(18438 y^{12} x^8-3792 y^{14} x^6+243y^{16} x^4+104 y^{18} x^2+7 y^{20})\alpha\beta}{x^3 y^3 (x^2-y^2)^4 (x^4-6 y^2 x^2+y^4)^3}\\
   & -\frac{64 (x^2+y^2)^4 (x^4-6 y^2 x^2+y^4) \beta ^3}{x^3 y^3 (x^2-y^2)^6}-\frac{256 (x^2+y^2)^4 (x^8+16 y^2 x^6-66 y^4 x^4+16 y^6 x^2+y^8) \alpha  \beta ^2}{x^3 y^3 (x^2-y^2)^4
   (x^4-6 y^2 x^2+y^4)^2}
\\
   &-\frac{8192 (x^2+y^2)^4 (x^8+10 y^2 x^6-30 y^4 x^4+10 y^6 x^2+y^8) \alpha ^2 \beta }{x y (x^2-y^2)^2 (x^4-6
   y^2 x^2+y^4)^4}\\
   &-\frac{1536 xy (5 y^{12}+158 x^2 y^{10}+395 x^4 y^8-540 x^6 y^6+395 x^8 y^4+158 x^{10} y^2+5 x^{12} ) \al \om^2}{(x^4-6 y^2
   x^2+y^4)^4}\\
   &+\frac{48 (x^{12}-4 y^2 x^{10}-17 y^4 x^8-152 y^6 x^6-17 y^8 x^4-4 y^{10} x^2+y^{12}) \beta  \om ^2}{x^3 y^3 (x^2-y^2)^4}
   \\
   &+\frac{8192 x y (x^2+y^2)^2 (x^8+6 y^2 x^6-22 y^4 x^4+6 y^6 x^2+y^8) \alpha ^2 \om ^2}{(x^4-6 y^2 x^2+y^4)^4}
\\
   &+\frac{64 (x^2+y^2)^2 (x^4-14 y^2 x^2+y^4) \beta ^2 \om^2}{x y (x^2-y^2)^4}
   -\frac{64 xy ( y^8-220 x^2 y^6+262 x^4 y^4-220 x^6 y^2+ x^8 ) \al \om^4}{(x^4-6 y^2 x^2+y^4)^2}
\\
 &+\frac{512 (x^2+y^2)^2 (x^8+8 y^2 x^6-50 y^4 x^4+8 y^6 x^2+y^8) \al \beta \om^2}{x y (x^6-7 y^2 x^4+7 y^4 x^2-y^6)^2}
\\
   &+\frac{4 (16 x^8-13 y^2 x^6+442 y^4 x^4-13 y^6 x^2+16 y^8) \beta \om^4}{x y (x^2-y^2)^2}-64 x y \om^4-64 x^3 y^3 \om^6 \bigg\}\\
   &+\bigg\{\pa_y^2,-\frac{15360 (3 x^{18}+249 y^2 x^{16}+1260 y^4 x^{14}-1068 y^6 x^{12}+666 y^8 x^{10}+318 y^{10} x^8) \alpha }{(x^4-6 y^2 x^2+y^4)^6}\\
   &-\frac{15360(-4932 y^{12} x^6+7140 y^{14} x^4+1467y^{16} x^2+17 y^{18})\alpha}{(x^4-6 y^2 x^2+y^4)^6}-\frac{120 (5 x^{18}-30 y^2 x^{16}+75 y^4 x^{14}) \beta }{x^6 y^6 (x^2-y^2)^6}
\\
   &-\frac{120(-87 y^6 x^{12}+333 y^8 x^{10}+165 y^{10} x^8+81 y^{12} x^6-45 y^{14} x^4+18 y^{16} x^2-3y^{18})\beta}{x^6 y^6 (x^2-y^2)^6}
\\
   &-\frac{2048 y^2 (303 x^{16}-1560 y^2 x^{14}+8628 y^4 x^{12}-16392 y^6 x^{10}+26538 y^8 x^8-17256 y^{10} x^6) \alpha ^2}{(x^4-6 y^2 x^2+y^4)^6}\\
   &-\frac{2048 y^2(7380 y^{12} x^4+264 y^{14} x^2+31y^{16})\alpha^2}{(x^4-6 y^2 x^2+y^4)^6}-\frac{8 (7 x^{18}-39 y^2 x^{16}+87 y^4 x^{14}-282 y^6 x^{12}) \beta ^2}{x^6 y^6 (x^2-y^2)^6}
\\
   &-\frac{8(1551 y^8 x^{10}-1242 y^{10} x^8+1281 y^{12} x^6-354 y^{14} x^4+18 y^{16} x^2-3 y^{18})\beta^2}{x^6 y^6 (x^2-y^2)^6}
\\
   &-\frac{128 (3 x^{20}-114 y^2 x^{18}+453 y^4 x^{16}-2160 y^6 x^{14}+7470 y^8 x^{12}-19604 y^{10} x^{10}) \alpha  \beta }{x^2 y^4 (x^2-y^2)^4 (x^4-6 y^2 x^2+y^4)^3}
\end{align*}
\begin{align*}
   &-\frac{128(21866 y^{12} x^8-17856 y^{14} x^6+7551
   y^{16} x^4-714 y^{18} x^2+33 y^{20})\alpha\beta}{x^2 y^4 (x^2-y^2)^4 (x^4-6 y^2 x^2+y^4)^3}\\
   &-\frac{8192 (x^2+y^2)^4 (x^{10}-3 y^2 x^8+34 y^4 x^6-46 y^6 x^4+45 y^8 x^2+y^{10}) \alpha ^3}{(x^4-6 y^2x^2+y^4)^6}\\
   &-\frac{2 (x^2+y^2)^4 (x^{10}-7 y^2 x^8+18 y^4 x^6-70 y^6 x^4+29 y^8 x^2-3 y^{10}) \beta ^3}{x^6 y^6 (x^2-y^2)^6}\\
   &-\frac{512 (x^2+y^2)^4 (3 x^{10}-13 y^2 x^8+86 y^4 x^6-162 y^6 x^4+119 y^8 x^2-y^{10}) \alpha ^2 \beta }{x^2 y^2 (x^2-y^2)^2 (x^4-6 y^2 x^2+y^4)^4}\\
   &-\frac{32 (x^2+y^2)^4 (3 x^{10}-17 y^2 x^8+70 y^4 x^6-186 y^6 x^4+103 y^8 x^2-5 y^{10}) \alpha  \beta ^2}{x^4 y^4
   (x^2-y^2)^4 (x^4-6 y^2 x^2+y^4)^2}\\
   &+\frac{256 (23 x^{14}+606 y^2 x^{12}-167 y^4 x^{10}+512 y^6 x^8-775 y^8 x^6+1030 y^{10} x^4-9x^2y^12-4 y^{14}) \alpha  \omega ^2}{(x^4-6 y^2x^2+y^4)^4}\\
   &+\frac{8 (9 x^{12}-32 y^2 x^{10}+91 y^4 x^8+112 y^6 x^6+367 y^8 x^4+16 y^{10} x^2+13 y^{12}) \beta \om^2}{x^2 y^4(x^2-y^2)^4}
\\
   &-\frac{512 (x^2+y^2)^2 (9x^{10}-215 y^2 x^8+354 y^4 x^6+506 y^6 x^4-315 y^8 x^2+13 y^{10}) \alpha ^2 \omega ^2}{(x^4-6 y^2
   x^2+y^4)^4}\\
   & +\frac{2 (x^2+y^2)^2 (9 x^{10}-51 y^2 x^8+142 y^4 x^6-26 y^6 x^4+25 y^8 x^2-3 y^{10}) \beta ^2 \omega ^2}{x^4 y^4(x^2-y^2)^4}\\
   &+\frac{128 (x^2+y^2)^2 (41 x^8-53 y^2 x^6-133 y^4 x^4-85 y^6 x^2-4 y^8) \alpha  \beta  \omega ^2}{(y^7-7 x^2 y^5+7 x^4y^3-x^6 y)^2}\\
 &-\frac{32 (24 x^{10}+5 y^2 x^8+76 y^4 x^6+630 y^6 x^4-156 y^8 x^2+13 y^{10}) \alpha  \omega ^4}{(x^4-6 y^2 x^2+y^4)^2}\\
   &-\frac{2 (19 x^{10}+10 y^2 x^8+212 y^4 x^6-171 y^6 x^4+99 y^8 x^2-9 y^{10}) \beta  \omega^4}{x^2 (x-y)^2 y^2 (x+y)^2}-96 y^2 \omega ^4+32 x^4 y^2 \omega ^6\bigg\} \\
   &+ \frac{10080 (x^{24}-8 y^2 x^{22}+28 y^4 x^{20}-56 y^6 x^{18}+103 y^8 x^{16}+832 y^{10} x^{14}+2296 y^{12} x^{12}) \beta }{x^8 y^8 \left(x^2-y^2\right)^8}\\
   &+\frac{10080(832 y^{14} x^{10}+103 y^{16}
   x^8-56 y^{18} x^6+28 y^{20} x^4-8 y^{22} x^2+y^{24})\beta}{x^8 y^8 \left(x^2-y^2\right)^8}\\
  &+\frac{768 \left(5 x^{32}-165 y^2 x^{30}+2490 y^4 x^{28}-16359 y^6 x^{26}+74944 y^8 x^{24}-219501 y^{10} x^{22}\right) \alpha  \beta }{x^6 (x-y)^6 y^6 (x+y)^6 \left(x^2-2 y x-y^2\right)^4 \left(x^2+2 y x-y^2\right)^4}
\\
   &+\frac{256 \left(x^2+y^2\right)^4 \left(x^{16}-31 y^2 x^{14}+190 y^4 x^{12}-865 y^6 x^{10}+1666 y^8 x^8-865 y^{10} x^6+\right) \alpha  \beta ^2}{x^6 y^6 \left(x^2-y^2\right)^6 \left(x^4-6 y^2 x^2+y^4\right)^2}\\
   &+\frac{256(190 y^{12} x^4-31 y^{14}x^2+y^{16})\alpha\beta^2}{x^6 y^6 \left(x^2-y^2\right)^6 \left(x^4-6 y^2 x^2+y^4\right)^2} -\frac{6144 \left(x^2+y^2\right)^2 \left(47 x^{16}+4184 y^2 x^{14}\right) \alpha  \omega ^2}{\left(x^2-2 y x-y^2\right)^6 \left(x^2+2 y x-y^2\right)^6}
\\
   &-\frac{6144(17892 y^4 x^{12}-37016 y^6 x^{10}+41818 y^8 x^8-37016 y^{10} x^6+17892 y^{12}
   x^4)\alpha\omega^2}{\left(x^2-2 y x-y^2\right)^6 \left(x^2+2 y x-y^2\right)^6}
\\
   &-\frac{6144(4184 y^{14} x^2+47 y^{16})\alpha\omega^2}{\left(x^2-2 y x-y^2\right)^6 \left(x^2+2 y x-y^2\right)^6}-\frac{48 \left(x^2+y^2\right)^2 \left(15 x^{16}-121 y^2 x^{14}+458 y^4 x^{12}\right) \beta  \omega ^2}{x^6 (x-y)^6 y^6 (x+y)^6}
\end{align*}
\begin{align*}
   &+\frac{768(584486 y^{12} x^{20}-1582599 y^{14} x^{18}+2641078
   y^{16} x^{16}-1582599 y^{18} x^{14})\alpha\beta}{x^6 (x-y)^6 y^6 (x+y)^6 \left(x^2-2 y x-y^2\right)^4 \left(x^2+2 y x-y^2\right)^4}\\
   &+\frac{768 \left(2490 x^4 y^{28}-16359 x^6 y^{26}+74944 x^8 y^{24}-219501 x^{10} y^{22}+584486 x^{12} y^{20}\right) \alpha  \beta }{x^6 (x-y)^6y^6 (x+y)^6 \left(x^2-2 y x-y^2\right)^4 \left(x^2+2 y x-y^2\right)^4}\\
   &+\frac{768(5 y^{32}-165 x^2 y^{30})\alpha\beta}{x^6 (x-y)^6y^6 (x+y)^6 \left(x^2-2 y x-y^2\right)^4 \left(x^2+2 y x-y^2\right)^4}\\
   &-\frac{65536 \left(x^2+y^2\right)^4 \left(x^8-76 y^2 x^6+166 y^4 x^4-76 y^6 x^2+y^8\right) \alpha ^3}{\left(x^4-6 y^2 x^2+y^4\right)^6}\\
   &+\frac{24576 \left(61 x^{16}+1608 y^2 x^{14}+11372 y^4 x^{12}-27400 y^6 x^{10}+44334 y^8 x^8-27400 y^{10} x^6\right) \alpha ^2}{\left(x^4-6 y^2 x^2+y^4\right)^6}\\
   &+\frac{24576(11372 y^{12} x^4+1608 y^{14} x^2+61
   y^{16})\alpha^2}{\left(x^4-6 y^2 x^2+y^4\right)^6}+\frac{12 \left(87 x^{24}-676 y^2 x^{22}+2246 y^4 x^{20}\right) \beta ^2}{x^8 y^8 \left(x^2-y^2\right)^8}\\
   &+\frac{12(-6932 y^6 x^{18}+24281 y^8 x^{16}+37304 y^{10} x^{14}+243732 y^{12} x^{12}+37304 y^{14}x^{10})\beta^2}{x^8 y^8 \left(x^2-y^2\right)^8}\\
   &+\frac{12(+24281 y^{16} x^8-6932 y^{18} x^6+2246 y^{20} x^4-676 y^{22} x^2+87 y^{24})\beta^2}{x^8 y^8 \left(x^2-y^2\right)^8}\\
   &+\frac{16 \left(x^2+y^2\right)^4 \left(x^{16}-11 y^2 x^{14}+70 y^4 x^{12}-245 y^6 x^{10}+626 y^8 x^8-245 y^{10} x^6+70 y^{12} x^4\right) \beta ^3}{x^8 y^8 \left(x^2-y^2\right)^8}\\
   &+\frac{16(-11 y^{14}
   x^2+y^{16})\beta^3}{x^8 y^8 \left(x^2-y^2\right)^8}-\frac{12288 \left(x^2+y^2\right)^4 \left(7 x^8-52 y^2 x^6+202 y^4 x^4-52 y^6 x^2+7 y^8\right) \alpha ^2 \beta }{x^2 y^2 \left(x^2-y^2\right)^2
   \left(x^4-6 y^2 x^2+y^4\right)^4} \\
   &  +\frac{65536 \left(x^2+y^2\right)^8 \alpha ^4}{\left(x^4-6 y^2 x^2+y^4\right)^6}+\frac{16384 \left(x^2+y^2\right)^8 \alpha ^3 \beta }{x^2 y^2 \left(x^2-y^2\right)^2 \left(x^4-6 y^2 x^2+y^4\right)^4}\\
     &-\frac{48(-999 y^6 x^{10}+5134 y^8 x^8-999 y^{10} x^6+458 y^{12} x^4-121 y^{14}
   x^2+15 y^{16})\beta\omega^2}{x^6 (x-y)^6 y^6 (x+y)^6}\\
   &+\frac{1536 \left(x^2+y^2\right)^8 \alpha ^2 \beta ^2}{x^4 y^4 \left(x^2-y^2\right)^4 \left(x^4-6 y^2 x^2+y^4\right)^2}+\frac{147456 x^2 (x-y)^2 y^2 (x+y)^2 \alpha  \om^4}{\left(x^2-2 y x-y^2\right)^2 \left(x^2+2 y x-y^2\right)^2}
\\
       &+\frac{64 \left(x^2+y^2\right)^8 \alpha  \beta ^3}{x^6 y^6 \left(x^2-y^2\right)^6}+\frac{512 \left(x^2+y^2\right)^2 \left(39 x^8+92 y^2 x^6+362 y^4 x^4+92 y^6 x^2+39 y^8\right) \alpha  \beta  \omega ^2}{x^2 y^2 \left(x^6-7 y^2x^4+7 y^4 x^2-y^6\right)^2}
\\
   & +\frac{\left(x^2+y^2\right)^8 \left(x^4-6 y^2 x^2+y^4\right)^2 \beta ^4}{x^8 y^8 \left(x^2-y^2\right)^8}-\frac{16 \left(x^2+y^2\right)^2 \left(3 x^{16}-23 y^2 x^{14}+54 y^4 x^{12}\right) \beta ^2 \omega ^2}{x^6 (x-y)^6 y^6 (x+y)^6}  \\
   &-\frac{16(-41 y^6 x^{10}+782 y^8 x^8-41 y^{10} x^6+54 y^{12} x^4-23 y^{14} x^2+3
   y^{16})\beta^2\omega^2}{x^6 (x-y)^6 y^6 (x+y)^6}\\
   &-\frac{4096 \left(x^2+y^2\right)^2 \left(x^{16}+568 y^2 x^{14}+124 y^4 x^{12}+3592 y^6 x^{10}-8314 y^8 x^8\right) \alpha ^2 \omega ^2}{\left(x^2-2 y x-y^2\right)^6 \left(x^2+2 y x-y^2\right)^6}\\
   &-\frac{4096(x^2+y^2)^2(3592 y^{10} x^6+124 y^{12} x^4+568 y^{14}
   x^2+y^{16})\alpha^2\omega^2}{\left(x^2-2 y x-y^2\right)^6 \left(x^2+2 y x-y^2\right)^6}
\\
   &+\frac{432 \left(x^8-2 y^2 x^6+18 y^4 x^4-2 y^6 x^2+y^8\right) \beta  \omega ^4}{x^2 (x-y)^2 y^2 (x+y)^2}-\frac{2304 \left(x^2+y^2\right)^6 \alpha  \beta ^2 \omega ^2}{x^2 y^2 \left(x^6-7 y^2 x^4+7 y^4 x^2-y^6\right)^2}
\end{align*}
\begin{align}
      \begin{split}
   &+\frac{3072 \left(x^2+y^2\right)^6 \left(3 x^8-58 y^2 x^6+158 y^4 x^4-58 y^6 x^2+3 y^8\right) \alpha ^2 \beta  \omega ^2}{x^2 y^2
   \left(x^2-y^2\right)^2 \left(x^4-6 y^2 x^2+y^4\right)^4}\\
   &+\frac{32768 \left(x^2+y^2\right)^6 \left(3 x^8-52 y^2 x^6+146 y^4 x^4-52 y^6 x^2+3 y^8\right) \alpha ^3 \omega ^2}{\left(x^4-6 y^2 x^2+y^4\right)^6}\\
   &-\frac{4 \left(x^2+y^2\right)^6 \left(3 x^8-34 y^2 x^6+110 y^4 x^4-34 y^6 x^2+3 y^8\right) \beta ^3 \omega ^2}{x^6 y^6 \left(x^2-y^2\right)^6}\\
   &+\frac{8192 x^2 y^2 \left(x^2-y^2\right)^2 \left(45 x^8-284 y^2 x^6+1198 y^4 x^4-284 y^6 x^2+45 y^8\right) \alpha ^2 \omega ^4}{\left(x^4-6 y^2x^2+y^4\right)^4}\\
   &+\frac{38 \left(x^{16}+2 y^2 x^{14}-18 y^4 x^{12}+198 y^6 x^{10}-110 y^8 x^8+198 y^{10} x^6-18 y^{12} x^4+2 y^{14} x^2+y^{16}\right) \beta ^2
   \om^4}{x^4 y^4 \left(x^2-y^2\right)^4}\\
   &-\frac{4 \left(x^2+y^2\right)^2 \left(3 x^8-34 y^2 x^6+110 y^4 x^4-34 y^6 x^2+3 y^8\right) \beta  \om^6}{x^2 y^2 \left(x^2-y^2\right)^2}\\
    &+\frac{512 \left(3 x^{16}-42 y^2 x^{14}+686 y^4 x^{12}-2702 y^6 x^{10}+4878 y^8 x^8-2702 y^{10} x^6+686 y^{12} x^4\right)\al \beta  \om^2}{x^2 y^2 \left(x^6-7 y^2 x^4+7 y^4 x^2-y^6\right)^2}\\
    &+\frac{512(-42 y^{14} x^2+3 y^{16})\al \beta  \om^2}{x^2 y^2 \left(x^6-7 y^2 x^4+7 y^4 x^2-y^6\right)^2}
\\
&+\frac{128 \left(x^2+y^2\right)^2 \left(3 x^8-52 y^2 x^6+146 y^4 x^4-52 y^6 x^2+3 y^8\right) \al \om^6}{\left(x^4-6 y^2 x^2+y^4\right)^2} +128 x^2 y^2 \om^6\ .
  \end{split}
   \end{align}
The algebraic form $y_8=\Psi_0^{-1}(\mathcal{Y}_8-C_8)\Psi_0$ after the change of variables $t=r^2$ and $u=r^8\sin^24\varphi$ is
\begin{align*}
  & y_8=256(t^4-u)\pa_t^8 -49152(t^4-u)t^2u\,\,\pa_t^6\pa_u^2-262144(t^4-u)tu^2\,\,\pa_t^5\pa_u^3 \\
  & +131072(19t^8-23t^4u+4u^2)u^2\pa_t^4\pa_u^4
  +25165824(t^4-u)t^3u^3\,\pa_t^3\pa_u^5 \\
  & -4194304(3t^8-31t^4u+28u^2)t^2u^3\,\,\pa_t^2\pa_u^6 -67108864(t^8-5t^4u+4u^2)tu^4\,\, \pa_t\pa_u^7 \\
  & +16777216(t^4-4u)^2(t^4-u)u^4\,\, \pa_u^8\\
  & -1024 \big[\om t^4-2(2a+1) t^3-\om u \big]\pa_t^7-24576 \big[(2 b+1) t^4-2(a+b+1) u\big]t^2\pa_t^6\pa_u\\
  & +49152  \big[3 \om  t^5-4 (3 a+2 b+6) t^4-3\om t u+2 (4 a+4 b+11) u\big]t u\pa_t^5\pa_u^2\\
  & +131072  \big[19 (2 b+3) t^8+5\om t^5 u-(58 a+46 b+111)t^4 u-5 \om tu^2+
  (8 a+8 b+29) u^2\big]u\pa_t^4\pa_u^3\\
 & -262144  \big[19 \om t^8-2(38 a+120 b+357) t^7-23\om t^4 u +4 (64 a+60 b+201)t^3 u+4\om u^2 \big]u^2\pa_t^3\pa_u^4\\
& -6291456  \big[3 (2b+5) t^8+6\om t^5 u-2 (15 a+31 b+122) t^4 u-6\om t u^2
+2 (28 a+28 b+121) u^2\big]t^2 u^2\pa_t^2\pa_u^5\\
&-4194304 \big[-3\om  t^9+4 (3a+14 b+55) t^8+31\om t^5 u-2 (84 a+140 b+681) t^4 u\\
 &-28 \om t u^2 + 56(4 a+4 b+21) u^2\big]t u^3\pa_t\pa_u^6\\
 &+33554432 \big[(2 b+7) t^{12}+\om t^9u -3 (2 a+6 b+29)t^8 u-5\om t^5 u^2 +(32 a+48 b+273) t^4 u^2 \\
 &+4 \om t u^3 -4 (8 a+8 b+49) u^3\big]u^3\pa_u^7\\
 &+512 \big[3 \om^2 t^4-14 (2 a +1)\om t^3+6 (2 a+1) (4 a+4 b+3) t^2
 -3 \om^2 u \big]\pa_t^6\\
 &+8192 \big[9 (2 b +1)\om t^5-4 (2 b+1) (9 a+2b+12) t^4-18 (a+b+1) \om t u
 +4 (8 a^2+(12 b+19) a\\
 &+4 b^2+23 b+15) u\big]t\pa_t^5\pa_u + 8192 \big[76 (2 b +1)(2b+3) t^8
 -18  \om^2 t^6 u+60(3a+2b+6) \om t^5 u\\
 &-(448 a^2+4 (348 b+635) a+3 (128 b^2 +576 b+631)) t^4u+18 \om^2 t^2u^2
 -30 (4 a+4 b+11) \om t u^2\\
   &+(112 a^2+8 (24 b+61) a+80 b^2+520 b+723) u^2\big]\pa_t^4\pa_u^2\\
    &-131072  \big[38 (2 b+3) \om  t^8-4 (2 b+3) (38 a+48 b+177) t^7+3  \om^2 t^5u-2 (58 a+46 b +111)  \om  t^4u \\
 &+8 (52 a^2+2 (64 b+163) a+48 b^2+294 b+387) t^3 u -3  \om ^2 t u^2
 + 2 (8 a+8 b+29)  \om u^2 \big]u\pa_t^3\pa_u^3\\
   &-131072 \big[72 (4 b^2+16 b+15) t^8-27  \om ^2 t^6u+6 (38 a+120 b+357)  \om  t^5u-4 (124 a^2\\
   &+25 (36 b+101) a+800 b^2+5590 b+9156) t^4 u +27  \om ^2 t^2u^2-12 (64 a+60 b +201) \om t u^2\\
   &+2 (1456 a^2+32 (105 b+358) a+1456 b^2+11384 b+20673) u^2\big] t^2 u\pa_t^2\pa_u^4\\
   &+2097152  \big[9 (2 b+5) \om  t^9
   -36 (2 b+5) (a+2 b+9) t^8+9  \om ^2 t^6u-6 (15 a+31 b+122) \om t^5u
   &+2 (112 a^2+2 (252 b+961) a+376 b^2+3290b+6939)t^4 u-9 \om ^2 t^2 u^2+6 (28 a+28 b+121)  \om  tu^2 \\
   &-2 (304 a^2+8 (84 b+361) a+304 b^2+2888 b+6609) u^2\big]tu^2\pa_t\pa_u^5
\end{align*}
\begin{align*}
   &+2097152 \big[12(2b+7)(2b+5) t^{12}+4 (3 a+14 b+55) \om t^9u-(32 a^2+4 (84 b+313) a+448 b^2\\
   &+3920 b+8313\om) u t^8+12 \om^2 t^6 u^2-2 (84 a+140 b+681)  \om t^5u^2+(496 a^2+8 (224 b+1051) a\\
   &+1232 b^2+12712 b+32223) u^2 t^4-12  \om^2 t^2u^3+56 (4 a+4 b+21)  \om  tu^3\\
   &-4(208 a^2+8 (56 b+289) a+208 b^2+2312 b+6321) u^3\big] u^2\pa_u^6\\
   &+1024 \big[-\om ^3 t^4+9 (2 a+1) \om ^2 t^3-9 (2 a+1) (4 a+4 b+3) \om  t^2+4 (2 a+1) (8 a^2 +2 (8 b+5) a\\
   &+8 b^2+10 b+3) t+\om^3u\big]\pa_t^5\\
   &+4096 \big[-18 (2 b+1) \om ^2 t^6 +20(2 b+1)(9 a+2(b+6)) \om t^5-(2 b+1)(448 a^2+4 (116 b+287) a\\
   &+16 b^2+320 b+609) t^4 +36 (a+b+1) \om^2
   t^2 u-20 (8 a^2+(12 b+19) a+4 b^2+23 b+15) \om t u\\
   &+2 (48 a^3+16 (7 b+11) a^2
   +(80 b^2+312 b+247) a+16 b^3+136 b^2+319 b+159)u\big]\pa_t^4\pa_u\\
   &-16384 \big[76 (2 b+1)(2b+3)\om t^8-8(38 a+24 b+117)(2 b +1)(2b+3) t^7
   -3 \om^3 t^6 u\\
   &+18 (4 a+2 b+7)\om^2 t^5 u-(448 a^2+4(348 b+635) a+3 (128 b^2+576 b+631)) \om t^4 u\\
   &+2(416 a^3+16(156 b+251)a^2+2(1232 b^2+5056 b+4813) a+384 b^3
   +3568 b^2+7856 b+5037) t^3 u\\
   &+3 \om^3 t^2 u^2-36 (a+b+3) \om ^2 t u^2 +(112 a^2+8(24 b+61) a
   +80 b^2+520 b+723)\om u^2 \big]\pa_t^3\pa_u^2\\
   &-65536 \big[24 (2b+5)(2b+3)(2b+1) t^{10}-54 (2 b+3)\om ^2 t^8 u
   +12(2 b+3)(38 a+48 b+177)\om t^7 u\\
   &-8 (2 b+3) (124 a^2+5 (72 b+235) a+4  (45 b^2+346 b+629)) t^6 u
   +\om ^3 t^5 u^2 \\
   &+9 (16 a+12 b+27)\om ^2 t^4 u^2 -24 (52 a^2+2 (64 b+163) a+48 b^2
   +294 b+387)\om t^3 u^2 \\
   &+4 (672 a^3+8 (364 b+901) a^2+4 (728 b^2+4160 b+5701) a+672 b^3+7136 b^2+22378 b +21351)t^2 u^2\\
   & -\om^3 t u^3 -9\om^2 u^3 \big]\pa_t^2\pa_u^3\\
  &+262144 [36 (2b+5)(2b+3)\om t^{10}-16 (9 a+10 b+53)(2b+5)(2b+3) t^9
  -4 \om ^3 t^8 u\\
  &+9 (6 a+20 b+59)\om ^2 t^7 u-2(124 a^2+25 (36 b+101) a+800 b^2+5590 b+9156)\om t^6 u\\
  &+4 (96 a^3+4 (280 b+757) a^2+(2176 b^2+14524 b+22919) a
  +960 b^3+11152 b^2+41048 b+47997) t^5 u \\
  &+2 \om ^3 t^4 u^2 -18 (10 a+10 b+33)\om ^2 t^3  u^2 +(1456 a^2
   +32 (105 b+358) a+1456 b^2+11384 b+20673) \om t^2  u^2 \\
   & -2 [1600 a^3+16 (380 b+1271) a^2 +(6080 b^2+44960 b+81732) a\\
   & +1600 b^3+20336 b^2+81684 b+104727] t u^2+2\om ^3  u^3 \big] u \pa_t\pa_u^4\\
   &+1048576  \big[8(2b+3)(2b+5)(2b+7) t^{12}+36 (2 b+5) (a+2 b+9) u \om  t^9-3 (2 b+5)\\
   & (32 a^2+4 (36 b+145) a+112 b^2+1040 b+2349)t^8 u-3 \om ^3 t^7  u^2+
   9(4 a+8 b+33)\om ^2 t^6 u^2 \\
   &-2 (112 a^2+2 (252 b+961) a+376 b^2+3290 b+6939)\om  t^5 u^2
   +2 [208 a^3+496 (3 b+11)  a^2\\
   &+(2416 b^2+20120 b+40965) a+3 (336 b^3+4648 b^2+20939 b+30720)]t^4 u^2
   +3 \om ^3 t^3 u^3
\end{align*}
\begin{align*}
   &-9(8 a+8 b+35)\om^2 t^2  u^3
   +2 (304 a^2+8 (84 b+361) a+304 b^2+2888 b+6609) \om  t  u^3\\
   &-2 [704 a^3+16(156 b +655) a^2+12 (208 b^2+1880 b+4241) a\\
   &+704 b^3+10480 b^2+50892 b+80847] u^3\big]\pa_u^5\\
   &+\big[256 \om^4 t^4 -256 \om^4 u -5120 (2a+1) \om^3 t^3 +
   128(664a^2+64a(9b+11)-b^2+289 b+227) \om^2 t^2 \\
   &-10240(2a+1)(8a^2+2a(8b+5)+8b^2+10b+3)\om t
   +1024(2a+1)(32a^3+48a^2(2b+1) \\
   &+2a(48b^2+48b+11)+32b^3+48b^2+22b+3)\big]\pa_t^4\\
   &-2048 \big[-12(2b+1)\om ^3 t^6+24(2b+1)(12a+2b+15) \om^2 t^5
   -4(2b+1)(448 a^2+4a(116 b+287)\\
   &+16 b^2+320 b+609)\om  t^4+8 (2 a+1)(2 b+1)(208 a^2+16a(26b+41)
   +208b^2 +656b+501)t^3 \\
   &+24(a+b+1)\om^3 t^2 u-(280a^2+8(36b+55)a+95b^2+625b+407)\om^2 t u\\
   &+8(48a^3+16 a^2(7b+11)+a(80b^2+312b+247)+16b^3+136b^2+319b+159)\om u
   \big]\pa_t^3\pa_u\\
   &+4096 \big[216 (2b +1)(2b+3) \om^2 t^8-48(38a+24 b+117)(2 b +1)(2b+3)
   \om  t^7+32(2b+1)(2b+3)\\
   &(124 a^2+5a(36 b+145)+56b^2+478b+1026)t^6+12(6 a-2 b+3)\om^3 t^5 u-
   (1064 a^2+8a(432 b+785)\\
   &+865 b^2+3671b+3931)\om^2 t^4 u+12(416a^3+16 a^2(156b+251)
   +2a(1232b^2+5056b+4813)\\
   &+384 b^3+3568 b^2+7856 b+5037)\om t^3 u-4(1792a^4+32 a^3(504 b+761)
   +16 a^2(1792 b^2+6836 b+6163)\\
   &+2a(8064 b^3+54544 b^2+113888 b+75225)+1792 b^4+24064 b^3+95776 b^2+142320 b\\
   &+70173)t^2 u+12 (2a+2b+1) \om^3 t u^2+2(88a^2+20a-b^2+109 b+254) \om^2 u^2\big]\pa_t^2\pa_u^2\\
   &-32768 \big[-48(2b+5)(2b+3)(2b+1)\om t^{10}+64 (3a+2b+13)(2b+5)(2b+3)(2b+1) t^9\\
   &+32(2 b+3)\om^3 t^8 u-72(2b+3)(6a+8b+29)\om ^2 t^7 u+16 (2b+3)
   [124a^2+5(72 b+235) a\\
   &+4 (45 b^2+346 b+629)] \om t^6 u -2 (4608 b^4+61568 b^3+288064 b^2+557184 b+\om^4 u \\
   &+1536 a^3 (2 b+3)+64 a^2 (224 b^2+1010 b+1011)+16 a (992 b^3+8480
   b^2+22186 b+17547)\\
   &+372096) t^5 u - 2(28 a+16 b+21)\om^3 t^4 u^2+[1064 a^2+8 (360 b+911) a+1153 b^2+6911 b\\
   &+8953] \om^2 t^3 u^2 -8(672 a^3+8 (364 b+901) a^2+4 (728 b^2+4160 b+5701) a+672 b^3+7136 b^2\\
   &+22378 b+21351) \om  t^2  u^2 +2[4096 a^4+256 a^3(100 b+239)+256 a^2(168 b^2+913 b+1227)\\
   &+32a(800 b^3+7304 b^2+21682 b+20973)+4096 b^4+u\om ^4+61184 b^3+313728 b^2+669264 b\\
   &+506232] t u^2 -2 (16 a+16 b+49)\om^3 u^3\big]\pa_t\pa_u^3\\
   &+32768 \big[32 (16 b^4+128 b^3+344 b^2+352 b+105) t^{12}+64(9a+10 b+53) (4 b^2+16 b+15)\om  t^9 u\\
   &+8 \om ^4t^8 u^2-8 (960 b^4+13440 b^3+65784 b^2+131136 b+192 a^2
   (4 b^2+16 b+15)+24 a (80 b^3\\
   &+676 b^2+1724 b+1335)+89190)t^8 u-8 (8 a+30 b+87)\om ^3 t^7  u^2+[472 a^2+80 (36 b+103) a\\
\end{align*}
\begin{align*}
   & +2495 b^2+18385 b+31043] \om ^2 t^6  u^2 -16 [96 a^3+
   4  a^2(280b+757)+a(2176 b^2+14524 b+22919) \\
   &+960 b^3+11152 b^2+41048 b+47997]\om t^5 u^2 -10\om ^4 t^4  u^3
   +2 (1024 a^4+128 a^3 (130 b+339) \\
   &+16  a^2(3216 b^2+20344 b+30887)+8 a(6240 b^3+67568 b^2+236602 b+266769) +14080 b^4\\
   &+227200b^3+1332928 b^2+3363192 b+3077145) t^4 u^2 +16 (13 a+15 b+48)
   \om^3 t^3 u^3\\
   &-48 (52 a^2+(120 b+413) a+52 b^2+413 b+756)\om^2 t^2  u^3+
   8 a^2(1600 a^3+16 (380 b+1271) \\
   &+a(6080 b^2+44960 b+81732)+1600 b^3+20336 b^2+81684 b+104727)
   \om t u^3 +2 \om^4 u^4 \\
   &-2 (10496 a^4+256  a^3(220 b+713)+32a^2(2864 b^2+20216 b+35941)
   +16a[3520 b^3+40432 b^2\\
   &+154588 b+196185]+10496 b^4 +182528 b^3+1150112 b^2+3138864 b+3138489) u^3\big]\pa_u^4\\
   &+\big[1024 (2 a+1)\om^4  t^3-256(184 a^2+8a(12 b+13)-b^2+49 b+47)
   \om^3 t^2+512 (368 a^3\\
   &+56 a^2(10 b+7)+2a(95 b^2+201b+99)-2 b^3+97 b^2+143 b+47) \om^2 t
   -2048(2a+1)\\
   &(32 a^3+48 a^2(2b+1)+a(96 b^2+96 b+22)+32b^3+48b^2+22b+3) \om\big]\pa_t^3\\
   &-1024 \big[-8 (18a-2b+15)(2b+1) \om^3 t^5+2(2b+1)[1064a^2+8a(144 b+353) +b^2+647 b\\
   & +1339]\om^2 t^4-24(2a+1)(2b+1)(208a^2+16a(26 b+41)+208 b^2+656 b+501)
   \om  t^3\\
   &+8(2a+1)(2b+1)[896a^3+16 a^2(168 b+229)+16a(168 b^2+458b+291)+896 b^3+3664 b^2\\
   &+4656 b+1809] t^2+(200 a^2-8a(12 b+25)-7(5 b^2-5 b-3))\om^3 t u
   -(704a^3+8 a^2(88b+105)\\
   &-8a(b^2-21b-7)-8b^3+285 b^2+1323 b+661) \om^2 u\big]\pa_t^2\pa_u\\
   &+4096 \big[-64(2b+1)(2b+3) \om^3 t^8+144 (6a+4b+19)(2b+1)(2b+3)
   \om ^2 t^7-32 (4 b^2+8 b+3)\\
   &(124 a^2+5a(36 b+145)+56 b^2+478b+1026) \om t^6+4(2 b+3)[1024 b^4+12544 b^3+52096 b^2\\
   &+81504 b+3 \om^4 u +1536 a^3(2 b+1)+64 a^2 (112 b^2+450 b+197)+
   16 a (320 b^3+2488 b^2\\
   &+5402 b+2119)+29232] t^5+(168 a^2+48a(14b+25)+97
   b^2+215 b+145) \om^3 t^4 u\\
   &-4(488 a^3+8 a^2(399 b+640)+a(3457 b^2+14087 b+13335)+579 b^3+5186 b^2+11158 b\\
   &+6992)\om^2 t^3 u+4 (1792 a^4+32 a^3(504 b+761)+16 a^2(1792 b^2+6836 b+6163)+2a(8064 b^3\\
   &+54544 b^2+113888b+75225)+1792 b^4+24064 b^3+95776 b^2+142320 b+70173) u \om  t^2\\
   &-8 u (1024 a^5+768 (16 b+23) a^4+128 (248 b^2+886
   b+765) a^3+128 (248 b^3+1496 b^2+2945 b\\
   &+1891) a^2+(12288 b^4+113408 b^3+376768 b^2+532416 b+3 \om^4 u +271632)
   a+1024 b^5+17664 b^4\\
   &+3 b \om^4 u +6 \om^4 u +97728 b^3+240560 b^2+268272 b+109332) t+2 (24 a^2+12 (16 b+39) a+81 b^2\\
   &+411b+469)\om^3 u^2 \big]\pa_t\pa_u^2
\end{align*}
\begin{align*}
   &-16384 \big[-64 (3 a+2 b+13) (8 b^3+36 b^2+46 b+15) \om t^9+16(2 b+3)
   (64 b^4+896 b^3+4196 b^2\\
   &+6892 b-\om^4 u +32a^2(4 b^2+12 b+5)+4 a(48 b^3+388 b^2+792 b+305)
   +2505) t^8\\
   &+32 (2b+3)(4a+6b+21)\om^3 t^7 u-2(2 b+3)(472a^2+16a(72b+245)+575 b^2
   +4705 b\\
   &+8843)\om^2 t^6 u+32(2b+3)[96a^3+4 a^2(112b+337)+a(496 b^2+3496 b+5849) +4 (36 b^3\\
   &+427 b^2+1610 b+1938)] \om t^5 u -2 u [6144 b^5+104960 b^4+686912 b^3+2143328 b^2-4 (5 \om^4 u \\
   &-793449) b-39 \om^4 u +2048 a^4 (2 b+3)+1024 a^3 (26 b^2
   +111 b+108)+32 a^2 (1472 b^3+11648 b^2\\
   &+28814b+21981)+4 a (7680 b^4+92544 b^3+401504 b^2+733984 b-3 \om^4 u +471048)+1771830] t^4\\
   &-(232 a^2+8 (104 b+259) a+385 b^2+2207 b+2761) \om^3 t^3 u^2 +3 (768 a^3+8 a^2(416 b+1029)\\
   &+32a(104 b^2+602 b+831)+768 b^3+8321 b^2+26335 b+25213)
   \om ^2 t^2  u^2-16 (512 a^4\\
   &+32 a^3(100 b+239)+32 a^2(168 b^2+913 b+1227)+4a(800 b^3+7304 b^2+21682 b+20973)\\
   &+512 b^4+7648 b^3+39216 b^2+83658 b+63279) \om t u^2
   +2 [5120 a^5+1024 a^4(41 b+94)\\
   &+128 a^3(784 b^2+4040 b+5255)+64 a^2(1568 b^3+13152 b^2+37202 b+35213)
   +4 a(10496 b^4\\
   &+129280 b^3+595232 b^2+1209232 b-\om^4 u+911949)+5120 b^5+96256 b^4-11 \om^4 u +672640 b^3\\
   &+2253440 b^2-4 b (\om^4 u -911745)+2284416]u^2 \big]\pa_u^3\\
   & +\big[128(88 a^2-16 a-b^2+b+11) \om^4 t^2 -256 a^2(464 a^3+8 (50 b+7)
   -2a(35 b^2+117 b+23)\\
   &-6 b^3-29 b^2+29 b+21)t \om ^3+256 (704 a^4+16 a^3(88 b+25)+8 a^2(87 b^2+35 b+10)\\
   &-2a(8 b^3+59 b^2-43 b-25)-8 b^4+2 b^3+93 b^2+67b+11) \om ^2\big]\pa_t^2\\
    &-1024 \big[-8 (2b +1)(2b+3) \om ^4 t^5-2 (2 b+1) (168 a^2+16 (14 b+33) a-31 b^2+7 b+121) \om ^3 t^4\\
   &+8 (2 b+1) (488 a^3+8 (133 b+241) a^2+(577 b^2+2391 b+2271) a+b^3+289 b^2+887 b +653) \om ^2 t^3\\
   & -8 (2 a+1) (2 b+1) (896 a^3+16 a^2(168 b+229)+16a(168 b^2+458 b+291)+896 b^3+3664 b^2\\
   &+4656 b+1809) \om  t^2+[16384 a^5(2b+1)+4096 a^4(32 b^2+58 b+21)+2048a^3 (96 b^3+308 b^2+292 b\\
   &+81) +8 a^2(16384 b^4+78848 b^3+128000 b^2+82944 b-3 u \om^4 +18304)+8a(4096 b^5+29696 b^4\\
   &+74752 b^3+82944 b^2+4 (3 \om^4 u +10304) b+21 \om^4 u +7488)+16384
   b^5+86016 b^4+73 \om^4 u \\
   &+165888 b^3 +11 b^2 (3 \om^4 u +13312)+3 b (37 \om^4 u +19968)+9216] t +[320a^3-8 a^2(24 b+71)\\
   &-24a(27 b^2+97 b+80)-136 b^3-803 b^2-1229 b-611] \om^3 u\big]\pa_t\pa_u\\
    &+256 \big[256 (2 b +1)(2b+3) \om ^4 t^8-256 (8 a+6 b+27) (2 b +1)(2b+3) \om ^3 t^7+32 (2 b +1)(2b+3)\\
   & (472a^2+16 (36 b+155) a+191 b^2+1681 b+3683) \om^2 t^6-512(2b +1)(2b+3) (96 a^3+4 a^2(56b+197)\\
   &+ô(160 b^2+1164 b+2119)+32 b^3+376 b^2+1440 b+1827) \om  t^5+[65536 b^6+1081344 b^5+6905856 b^4
\end{align*}
\begin{align*}
   &+21886976 b^3+(35698944-1443 \om^4 u) b^2+(27362304-4125 \om^4 u) b -3184 \om^4 u \\
   &+65536 a^4 (2b+1)(2b+8192 a^3 (104 b^3+524 b^2+710
   b+237)+16 a^2 (61440 b^4+505856 b^3\\
   &+1417984 b^2+1499136 b-19 \om^4 u +454464)+16 a (28672 b^5+337920 b^4+1498368 b^3\\
   &+3033728 b^2-48 (3 \om^4 u -55636) b-197 \om^4 u +743904)+7093440] t^4+32 (72 a^3
   +24 (29b+46)a^2
\\
   &+(993 b^2+3975 b+3709) a+195 b^3+1618 b^2+3302 b+1955) \om^3 t^3 u-24 (1024 a^4
   +96 (96 b
\\
   &+143) a^3+8 (2048 b^2+7804 b+7051) a^2+4 (2304 b^3+15873 b^2+33471 b+22213) a+1024 b^4\\
   &+14092 b^3+56707 b^2+84429 b+41523)\om ^2 t^2u+256 (256 a^5+192 (16 b+23) a^4
   +32 (248 b^2
\\
   &+886 b+765) a^3+32 (248 b^3+1496 b^2+2945 b+1891) a^2+4 (768 b^4+7088 b^3+23548 b^2
\\
   &+33276 b+16977) a+256 b^5+4416 b^4+24432 b^3+60140 b^2+67068 b+27333)\om t u\\
   &+(-65536 a^6-32768(30 b+41) a^5-24576 (136 b^2+452 b+373) a^4-2048 (2368 b^3+12992 b^2\\
   &+24036 b+14839) a^3-16 (208896 b^4+1662976
   b^3+5007360 b^2+6692864 b-47 \om^4 u \\
   &+3328016) a^2-16 (61440 b^5+694272 b^4+3076608 b^3+6692480 b^2+(7104112-48\om^4 u)b
\\
   &-61 \om^4 u +2932896) a-65536 b^6-1343488 b^5-9166848 b^4+411 b^2 \om^4 u
   +1317 b \om^4 u +1544 \om^4 u
\\
   &-30384128b^3-53208320 b^2-46851456 b-16147584)u\big]\pa_u^2
\\
   &+\big[256 (112 a^3+8 (6 b-7) a^2-2 (33 b^2+87 b+29) a-2 b^3-31 b^2-17 b-1) \om^4 t
\\
   &-256 (4 a+4 b+1) (112 a^3+8 (6 b-7)a^2-2 (33 b^2+87 b+29) a-2 b^3-31 b^2-17 b-1)
   \om^3\big]\pa_t
\\
   &+\big[-128(2b+1)(304 a^2+16 (48 b+53) a+163 b^2+413 b+496) \om^4 t^4\\
 &+256(496 a^3+16 a^2(47b+27)+a(411 b^2+181 b-56)+155 b^3+318 b^2+399 b+246)\om^4 u
\\
   &+4096(2b+1)(72a^3+8 a^2(29b+51)+a(161b^2+663 b+613)+b^3+81b^2+231b+152) \om^3 t^3
\\
   &-3072(2b+1)(1024 a^4+96 a^3(32b+47)+8 a^2(384b^2+1236b+943)+4a(256 b^3+1473 b^2\\
   &+2495b+1237)+516 b^3+2175 b^2+2737 b+1047) \om^2 t^2 +262144 (2a+1)
   (2 b+1)[16 a^4
\\
   &+4a^3(16b+19)+4 a^2(24b^2+57b+31)+a(64 b^3+228 b^2+248 b+81)+(4b+3)(2b+3)(2b+1)\\
   &(b+2)] \om t -256 [32768 a^6(2b+1)+16384 a^5(10b+11)(2b+1)+4096 a^4(160 b^3+440 b^2\\
   &+366b+93)+ 1024a^3(64b^4+2560b^3+3512b^2+1954b+379)+ 128a^2(2b+1)(1280b^4+6400b^3\\
   &+10848b^2+7232b+1577) +128a(2b+1)(256b^5+1920b^4+4896b^3+5368b^2+2509b+399)
\\
   &+32(2b+1)^2(4b+1)(4b+3)(16b^2+56b+51)]\big]\pa_u \ ,
\end{align*}
where $C_8$ is the lowest eigenvalue of $\mathcal{Y}_8$ (see (\ref{C8})).

\end{document}